# Comparison of optical spectra between asteroids Ryugu and Bennu:
# II. High-precision analysis for space weathering trends

K. Yumoto[a, *], E. Tatsumi[b], T. Kouyama[c], D. R. Golish[d], Y. Cho[a], T. Morota[a], S. Kameda[e, b], H. Sato[b], B. Rizk[d], D. N. DellaGiustina[d], Y. Yokota[b], H. Suzuki[f], J. de León[g, h], H. Campins[i], J. Licandro[g, h], M. Popescu[j], J. L. Rizos[k, l], R. Honda[m, †], M. Yamada[n], N. Sakatani[b], C. Honda[o], M. Matsuoka[p], M. Hayakawa[b], H. Sawada[b], K. Ogawa[q], Y. Yamamoto[b], D. S. Lauretta[d], S. Sugita[a, n, r]

[a]Department of Earth and Planetary Science, The University of Tokyo, Bunkyo, Tokyo, Japan.
[b]Institute of Space and Astronautical Science, Japan Aerospace Exploration Agency, Sagamihara, Kanagawa, Japan.
[c]Artificial Intelligence Research Center, National Institute of Advanced Industrial Science and Technology, Koto, Tokyo, Japan.
[d]Lunar and Planetary Laboratory, University of Arizona, Tucson, AZ, USA
[e]Department of Physics, Rikkyo University, Toshima, Tokyo, Japan.
[f]Department of Physics, Meiji University, Kawasaki, Kanagawa, Japan.
[g]Instituto de Astrofísica de Canarias (IAC), University of La Laguna, La Laguna, Tenerife, Spain.
[h]Department of Astrophysics, University of La Laguna, La Laguna, Tenerife, Spain.
[i]Department of Physics, University of Central Florida, Orlando, FL, USA.
[j]Astronomical Institute of the Romanian Academy, Bucharest, Romania.
[k]Instituto de Astrofísica de Andalcía (IAA) – CSIC, Granada, Andalcía, Spain.
[l]University of Maryland, College Park, MD, USA.
[m]Center for Data Science, Ehime University, Matsuyama, Ehime, Japan.
[n]Planetary Exploration Research Center (PERC), Chiba Institute of Technology, Narashino, Chiba, Japan.
[o]The University of Aizu, Aizu-Wakamatsu, Fukushima, Japan.
[p]Geological Survey of Japan (GSJ), National Institute of Advanced Industrial Science and Technology, Tsukuba, Ibaraki, Japan
[q]JAXA Space Exploration Center, JAXA, Sagamihara, Japan.
[r]Research Center of Early Universe, Graduate School of Science, The University of Tokyo, Tokyo, Japan.
†Deceased

*Corresponding author: Koki Yumoto (kyyumoto@gmail.com)

**Key points**
- The spectra of freshest craters on Ryugu and Bennu are indistinguishable.
- Spectral distributions of craters on Ryugu and Bennu form a common trend line.
- Ryugu and Bennu experienced divergent spectral change from similar initial spectra.
- Space weathering may have expanded the spectral slope variation of C-type asteroids.

**Abstract**

Various natural effects gradually alter the surfaces of asteroids exposed to the space environment. These processes are collectively known as space weathering. The influence of space weathering on the observed spectra of C-complex asteroids remains uncertain. This has long hindered our understanding of their composition and evolution through ground-based telescope observations. Proximity observations of (162173) Ryugu by the telescopic Optical Navigation Camera (ONC-T) onboard Hayabusa2 and that of (101955) Bennu by MapCam onboard Origins, Spectral Interpretation, Resource Identification, and Security-Regolith Explorer (OSIRIS-REx) found opposite spectral trends of space weathering; Ryugu darkened and reddened while Bennu brightened and blued. How the spectra of Ryugu and Bennu evolved relative to each other would place an important constraint for understanding their mutual relationship and differences in their origins and evolutions. In this study, we compared the space weathering trends on Ryugu and Bennu by applying the results of cross calibration between ONC-T and MapCam obtained in our companion paper. We show that the average Bennu surface is brighter by $18.0 ± 1.5\%$ at $v$ band (550 nm) and bluer by $0.18 ± 0.03$ μm$^{-1}$ (in the 480–850 nm spectral slope) than Ryugu. The spectral slopes of surface materials are more uniform on Bennu than on Ryugu at spatial scales larger than ~1 m, but Bennu is more heterogeneous at scales below ~1 m. This suggests that lateral mixing of surface materials due to resurfacing processes may have been more efficient on Bennu. The reflectance−spectral slope distributions of craters on Ryugu and Bennu appeared to follow two parallel trend lines with an offset before cross calibration, but they converged to a single straight trend without a bend after cross calibration. We show that the spectra of the freshest craters on Ryugu and Bennu are indistinguishable within the uncertainty of cross calibration. These results suggest that Ryugu and Bennu initially had similar spectra before space weathering and that they evolved in completely opposite directions along the same trend line, subsequently evolving into asteroids with different disk-averaged spectra. These findings further suggest that space weathering likely expanded the spectral slope variation of C-complex asteroids, implying that they may have formed from materials with more uniform spectral slopes.

1. **Introduction**

The spectral diversity observed among asteroids in the visible wavelengths (e.g., Bus & Binzel, 2002b; DeMeo et al., 2015; Hasselmann et al., 2015; Popescu et al., 2019) may represent compositional variations among and within primordial bodies in the solar system. The spectra of asteroids potentially provide insights into parent-body processes on these primordial bodies. However, obtaining such insights has been challenging because asteroid surfaces undergo alterations due to exposure to the space environment. Such processes may alter the observed spectra and superimpose its own signature on the intrinsic spectra of asteroid materials. Here, we use the term "space weathering" in a broad sense to encompass all possible alterations (both physical and chemical) that can occur on the exposed surfaces of asteroids (Pieters & Noble, 2016). Processes involved in space weathering include solar wind irradiation, micrometeorite bombardment, and the physical modification of surface materials by impacts and thermal effects.

The spectral changes induced by space weathering may be more complicated on C-complex asteroids than on S-complex asteroids and the Moon (e.g., Brunetto et al., 2015; Lantz et al., 2017). Laboratory experiments and telescopic observations of asteroids collectively show that the optical spectra of surface materials dominated by anhydrous silicates, such as those found on S-complex asteroids and the Moon, become darker and redder primarily due to the production of nanometer-sized metallic iron ($npFe^0$; e.g., Chapman, 2004; Hiroi et al., 2006). In contrast, laboratory experiments on carbonaceous chondrites show that the qualitative impact of space weathering—whether it results in brightening or darkening, as well as reddening or bluing—can be influenced by various conditions (Lantz et al., 2017; Clark et al., 2023). These conditions include the chemical/physical state of the asteroid material and the weathering agents responsible for the spectral modification (e.g., solar wind irradiation vs. micrometeorite bombardment). For instance, irradiation by solar wind-like ions causes CM chondrites to brighten (Lantz et al., 2017), whereas bombardment by micrometeorites simulated using pulsed lasers causes them to darken (Matsuoka et al., 2015). In addition, researchers have searched for evidence of space weathering effects using asteroid spectra observed by ground-based telescopes and found ambivalent trends that depend on how they estimate the age of asteroids. They observe a trend suggesting that asteroids redden over time when their age is estimated from an asteroid's collisional lifetime and orbit (Lazzarin et al., 2006), while a bluing trend is found when their age is estimated from the breakup ages of asteroid families (Nesvorný et al., 2005). More recently, different space weathering trends have been identified for different families by analyzing the size–color relationship among the family members (Thomas et al., 2021).

Remote sensing of two near-Earth carbonaceous rubble-pile asteroids, (162173) Ryugu and (101955) Bennu, by the sample return missions Hayabusa2 and OSIRIS-REx, respectively, has revealed evidence of space weathering effects on these asteroids (Sugita et al., 2019; Lauretta et al., 2019a; DellaGiustina et al., 2020; Morota et al., 2020; Tatsumi et al., 2021; Lauretta et al.,

2022). Ryugu and Bennu are asteroids with different spectral slopes in the visible wavelengths; Ryugu has a slightly red slope (Cb-type spectrum) while Bennu has a blue slope (B-type spectrum). Multi-band CCD imagers, the telescopic Optical Navigation Camera (ONC-T) aboard Hayabusa2 and the MapCam medium-field imager aboard OSIRIS-REx, observed high-resolution spectral images of the asteroid surface at visible wavelengths. Space weathering effects can be inferred from the spectra of craters observed by ONC-T and MapCam. This is because smaller craters are typically younger than larger ones, as they are more susceptible to being erased from the asteroid surface and are produced more frequently. Serendipitously, these two asteroids manifest opposite trends of space weathering. The spectra of small craters on Ryugu are bluer than the surrounding terrains, suggesting that the surface has reddened over time (Sugita et al., 2019; Morota et al., 2020). In contrast, on Bennu, the spectra of small craters are redder than the surrounding areas, indicating that the surface has become bluer (DellaGiustina et al., 2020; Lauretta et al., 2022), although subsurface stratigraphy may also contribute to this trend (Bierhaus et al., 2023).

Determining the underlying cause of the opposing space weathering trends may not be straightforward, even after samples have been returned. For instance, one challenge may arise from differences in physical properties, such as porosity and inter-grain contacts, between materials on the asteroid surface under microgravity conditions and those in laboratories on Earth, where the returned samples are analyzed. Laboratory experiments have shown that carbonaceous chondrites with different porosities (e.g., chips vs. pellets) exhibit contrasting spectral effects of space weathering (Nakamura et al., 2019; Clark et al., 2023)—a phenomenon not observed for anhydrous silicates (Marchi et al., 2005). Such a result suggests that space weathering effects on highly porous materials on the actual surfaces of Ryugu and Bennu (Lauretta et al., 2022; Walsh et al., 2022; Okada et al., 2020) may differ from those simulated in Earth-based laboratories. Hence, analyses of on-site remote sensing data may still retain critically important information that cannot be learned solely from sample analyses.

Quantitative comparison of the space weathering trends observed on Ryugu and Bennu is particularly important. For instance, DellaGiustina et al. (2020) and Clark et al. (2023) suggested that the spectral shapes of fresh craters may be similar between Ryugu and Bennu. This finding implies that Ryugu and Bennu may have had similar initial spectra before their opposite space weathering effects. However, previous studies could not ascertain whether the fresh craters on Ryugu and Bennu are also comparable in their absolute reflectances. This was because there was a large systematic error between the absolute reflectance data of Ryugu and Bennu, as ONC-T and MapCam used different targets for radiometric calibration (Yumoto et al., 2024).

To address this issue, we compare the space weathering trends on Ryugu and Bennu using the cross-calibrated data (i.e., data corrected for the calibration differences between ONC-

T and MapCam) developed in our companion paper (Yumoto et al., 2024). Such an analysis will enable us to quantitatively compare the opposing space weathering trends on the same absolute scale and determine whether these two asteroids evolved from a single initial spectrum or from two different initial spectra. We gain insights into the spectra of fresher materials than those obtained in previous studies by investigating the spectra of smaller craters identified in the latest crater counts by Takaki et al. (2022) and Bierhaus et al. (2022).

In the following sections, we first outline the methods for data preparation in section 2. Using the cross-calibrated multi-band images, we compare the optical spectral properties of Ryugu and Bennu on a global scale in section 3.1 and for craters in section 3.2. Based on these results, we discuss the implications for the opposite space weathering trends on Ryugu and Bennu in section 4. Finally, in section 5, we discuss the general implications for the spectral variation of C-complex asteroids.

## 2. Methods

### 2.1 Data selection and preprocessing for comparison of Ryugu and Bennu

We selected images of Ryugu and Bennu taken at similar geometries and spatial resolutions to conduct their unbiased comparison (Table 1). Selecting images taken at similar solar phase angle is especially important to minimize the error in photometric correction. For Bennu, we selected MapCam images observed with a phase angle of 9º taken during the hyperbolic flyby on September 26, 2019 (DellaGiustina et al., 2020). This dataset has a full global coverage with a high spatial resolution of 0.3 m/pix, allowing us to evaluate the spectra of meter-scale craters. ONC-T conducted high-resolution circum-equatorial mapping of Ryugu with a phase angle of 13º and a resolution of 0.3 m/pixel after the deployment of the Mobile Asteroid Surface Scout (MASCOT) lander on October 3–4, 2018. Since the coverage of these images is limited to the equatorial range (15ºN–30ºS), we also used lower-resolution (2 m/pix) images of Ryugu taken on January 31, 2019 with a phase angle of 10º to evaluate the spectra at higher latitudes.

**Table 1.** Images used for analyses and their observation geometries.

|       | Observation Date | Solar phase angle (°) † | Resolution (m/pix) † | Latitude range |
|-------|------------------|-------------------------|----------------------|----------------|
| Ryugu | Oct 3–4, 2018    | 12.5 ± 0.4              | 0.35 ± 0.05          | 15ºN – 30ºS    |
|       | Jan 31, 2019     | 9.9 ± 0.1               | 2.16 ± 0.04          | 90ºN – 90ºS    |
| Bennu | Sep 26, 2019     | 8.9 ± 0.5               | 0.26 ± 0.02          | 90ºN – 90ºS    |

†Ranges show the standard deviation within the dataset.

The data pre-processing proceed as follows. First, we prepared images of Ryugu and Bennu calibrated to radiance factors ($r_{obs,n}$) provided by the Planetary Data System; hereafter radiance factor is simply referred to as reflectance. These images are labelled "iofL2" for MapCam data (Rizk et al., 2019) and "L2d" for ONC-T data (Sugita et al., 2022). The subscript $n$ indicates the pertinent bands. ONC-T observed Ryugu at seven bands: $n = ul$ (398 nm), $b$ (480 nm), $v$ (549 nm), $Na$ (590 nm), $w$ (700 nm), $x$ (857 nm), and $p$ (945 nm) bands. MapCam observed Bennu at four bands: $n = b'$ (473 nm), $v$ (550 nm), $w$ (698 nm), and $x$ (847 nm) bands. Second, we performed band-to-band image co-registration to calculate the spectra at each pixel. Third, we multiplied the $r_{obs,n}$ of MapCam data by the imager-to-imager bias correction factor ($F_n$) derived from cross calibration (Table 5 in Yumoto et al., 2024) and obtain the cross-calibrated $r_{obs,n}$ ($r'_{obs,n}$):

$$r'^{MapCam}_{obs,n} = F_n \cdot r^{MapCam}_{obs,n}; \quad r'^{ONC}_{obs,n} = r^{ONC}_{obs,n} \qquad (1)$$

The multiplication by $F_n$ corrects the bias of MapCam to ONC-T caused by their differences in targets used for radiometric calibration and solar irradiance models used for data reduction. This correction essentially upscales the reflectance of Bennu by 13%, while it only changes the band ratios by <1.5%. Lastly, we photometrically corrected each pixel observed with incidence, emission, phase angles of $(i, e, \alpha)$ to a standard condition of $(i_{std}, e_{std}, \alpha_{std})=(10º, 0º, 10º)$ using the following equation:

$$r'_{norm,n}(i_{std}, e_{std}, \alpha_{std}) = r'_{obs,n}(i, e, \alpha) \cdot \frac{r_{model,n}(i_{std},e_{std},\alpha_{std})}{r_{model,n}(i,e,\alpha)}. \qquad (2)$$

Here $r_{model,n}$ is the reflectance modelled by disk-resolved photometric functions of the two asteroids (Tatsumi et al., 2020; Golish et al., 2020) and $r'_{norm,n}$ is the photometrically-corrected reflectance. We computed the observed $(i, e, \alpha)$ condition for each pixel using the asteroid shape models: the laser-altimeter-based shape model with 40 cm facet size (version 21) for Bennu and the stereo-photoclinometry-based shape model with 1 m facet size for Ryugu (version 23 Mar 2020). Since $|\alpha - \alpha_{std}|$ was <3º, the photometric correction factor $\left|\frac{r_{model,n}(\alpha_{std},0º,\alpha_{std})}{r_{model,n}(\alpha,0º,\alpha)} - 1\right|$ was only ≲0.05.

We produced equirectangular projections of $r'_{norm,n}$ using the latitude and longitude of each pixel calculated from the asteroid shape models. The latitude–longitude resolutions of these projections were determined to ensure that the sampling frequency at the equator matches the spatial resolution of the observed image (Table 1). We excluded pixels with $i$>70º or $e$>70º due to the large error in photometric correction. In addition, we set 0.01 of reflectance as the threshold for shadows; we excluded pixels below this threshold from our analysis.

Using these equirectangular projections, we compare the $r'_{norm,n}$ of Ryugu and Bennu at $n = b, v, w$, and $x$ bands, which are the only four bands shared between ONC-T and MapCam;

see section 2 in Yumoto et al. (2024) for comparability of data at these four bands. All analyses conducted in this study weight $r'_{norm,n}$ at each latitude–longitude grid point with the cosine of their latitudes to correct for the projected area.

## 2.2 Spectral analyses

We compared the $r'_{norm,n}$ of Ryugu and Bennu over the entire asteroid surface by calculating the global averages, latitudinal trend, and spatial variability (section 3.1). Space weathering effects may be inferred from latitudinal trends because global mass movements expose materials with different exposure ages at various latitudes (Sugita et al., 2019; DellaGiustina et al., 2020). For instance, on Bennu, global mass movements likely transported materials up to 10 meters thick from the mid-latitudes to the equator over the past several hundred thousand years, resulting in fresher materials being exposed at the mid-latitudes compared to the equatorial region (Jawin et al., 2020).

By averaging $r'_{norm,n}$ within each impact crater, we investigated its relationship with crater size to determine the space weathering trends (section 3.2). This is because crater size serves as an indicator of how long its floor material has likely been exposed to space. Fresher materials are more likely to be exposed at the floors of smaller craters because smaller craters have shorter retention ages; they are erased from the surface over shorter timescales by resurfacing processes (Takaki et al., 2022; Bierhaus et al., 2022). The retention age of craters with a diameter of 100 m is $10^6$–$10^7$ years, while that of craters with a diameter of 10 m is $10^3$–$10^6$ years (Takaki et al., 2022). Additionally, because the formation of smaller craters occurs more frequently due to the power-law increase in the number of smaller impactors, the freshest materials are likely found in the smallest craters (Bottke Jr et al., 2005).

We referred to Takaki et al. (2022) and Bierhaus et al. (2022) for the locations and sizes of craters. Takaki et al. (2022) identified 322 craters with diameters 2–308 m within the 40ºS to 40ºN region. Bierhaus et al. (2022) identified 1560 craters with diameters 0.8–215 m over the entire asteroid surface. We narrowed down the number of craters for analysis to 123 for Ryugu and 564 for Bennu based on the following criteria. We excluded craters situated at latitudes >60º because they were observed at oblique geometries and have low photometric correction accuracy. We restricted our analysis to craters larger than 4 m to ensure that each crater spectrum was obtained by averaging over >100 pixels. Additionally, we excluded ambiguous craters assigned with a confidence level of "4" in Takaki et al. (2022) to avoid possible false-positive identifications of craters. One possible false positive crater (identification number 1040 in Bierhaus et al., 2022), which is located immediately west of the Hokioi crater, was also excluded from analysis; later high-resolution imaging supports that the region is part of the ejecta blanket from the Hokioi crater (Fig. 4 in Barnouin et al., 2022).

Our analyses primarily focus on two spectral parameters, which are *v*-band reflectance ($r'_{norm,v}$) and *b*-to-*x* band spectral slope (i.e., the slope fitted over the *b*, *v*, *w*, and *x*-band (480–850 nm) spectrum of $r'_{norm,n}$ normalized at *v* band). This is because these two parameters exhibit the largest variations across the asteroid surface in the visible wavelengths (Sugita et al., 2019; DellaGiustina et al., 2019). These two parameters also account for most of the spectral variation among craters (Fig. S1b and c). Nevertheless, since crater spectra also show slight variations in the absorption depths at the *v* band (Fig. S1d; DellaGiustina et al., 2020), we present the results based on the *b*/*v* band ratio and *v*-to-*x* band spectral slope in the supplementary materials (Fig. S2).

## 3. Results and discussion: Post cross-calibration comparison of Ryugu and Bennu

We analyze the spectra across the entire asteroid surface (section 3.1) and among craters (section 3.2) for precise post cross-calibration comparison of the two asteroids and to search for space weathering trends. Our results show that the space weathering trends inferred from these two analyses are qualitatively similar. However, more detailed analyses were performed on the trend obtained from craters because the trend obtained from the entire surface appears to be muted by lateral mixing due to resurfacing processes.

### 3.1 Comparison of spectra across the entire asteroid surface

The equirectangular projection of $r'_{norm,v}$ (Fig. 1) and that of the *b*-to-*x* band spectral slope (Fig. 2) suggest that Bennu is brighter and bluer than Ryugu, but certain surface features exhibit similar spectra. These maps also suggest possible difference in spatial distributions. The spectral slope heterogeneity on Bennu appears to be dominated by variation among individual boulders larger than sub-meter size. In contrast, the heterogeneity on Ryugu has a clearer contribution from latitudinal and large-scale geologic units (e.g., large craters, troughs, and equatorial ridge), which are units mainly comprised of sub-pixel size (i.e., <0.3 m) regolith materials (Sugita et al., 2019; Morota et al., 2020). We quantitatively evaluate such findings in the following sections 3.1.1–3.1.3.

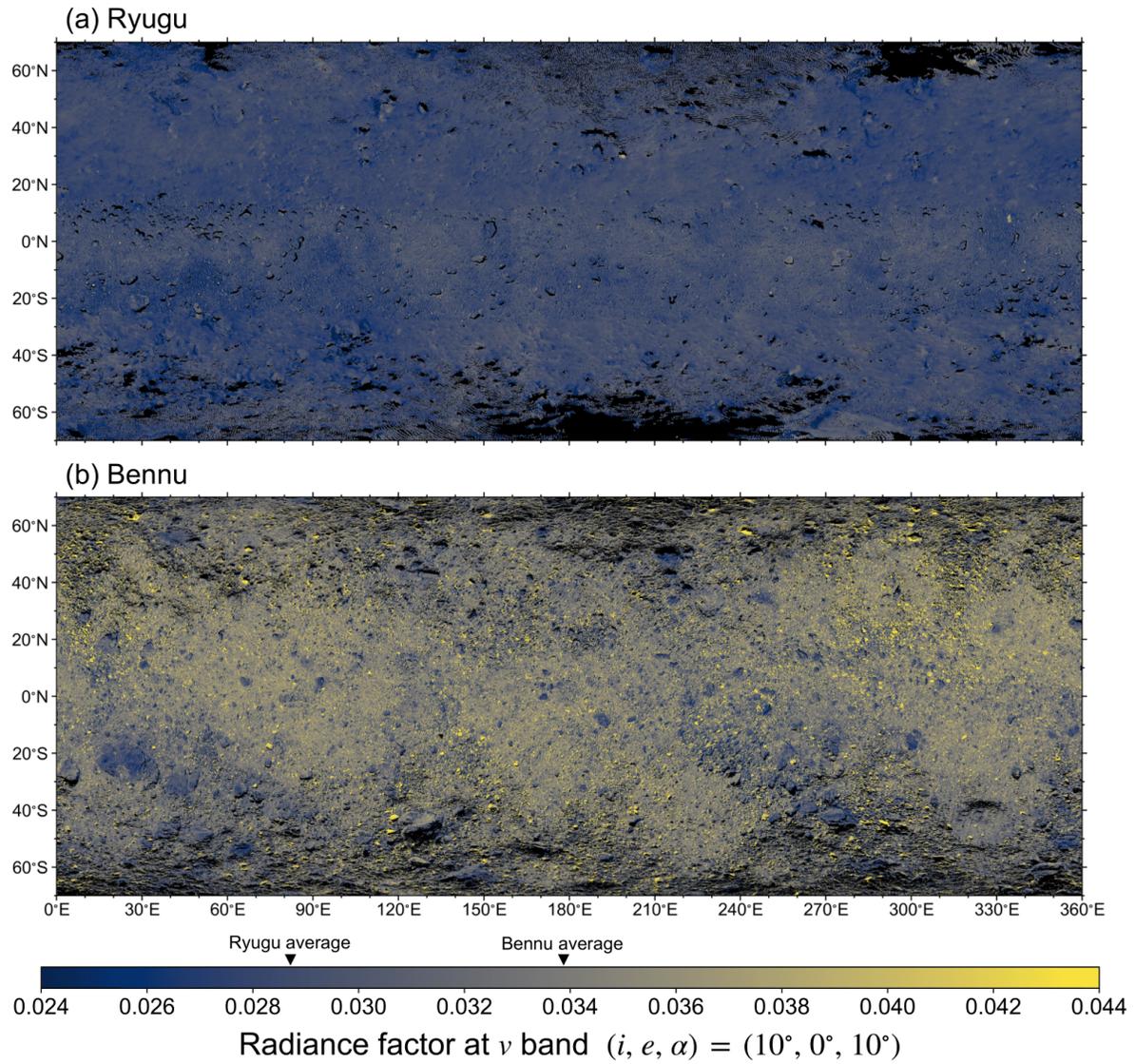

**Figure 1.** Equirectangular projections of *v*-band reflectance ($r'_{norm,v}$) of **(a)** Ryugu and **(b)** Bennu.

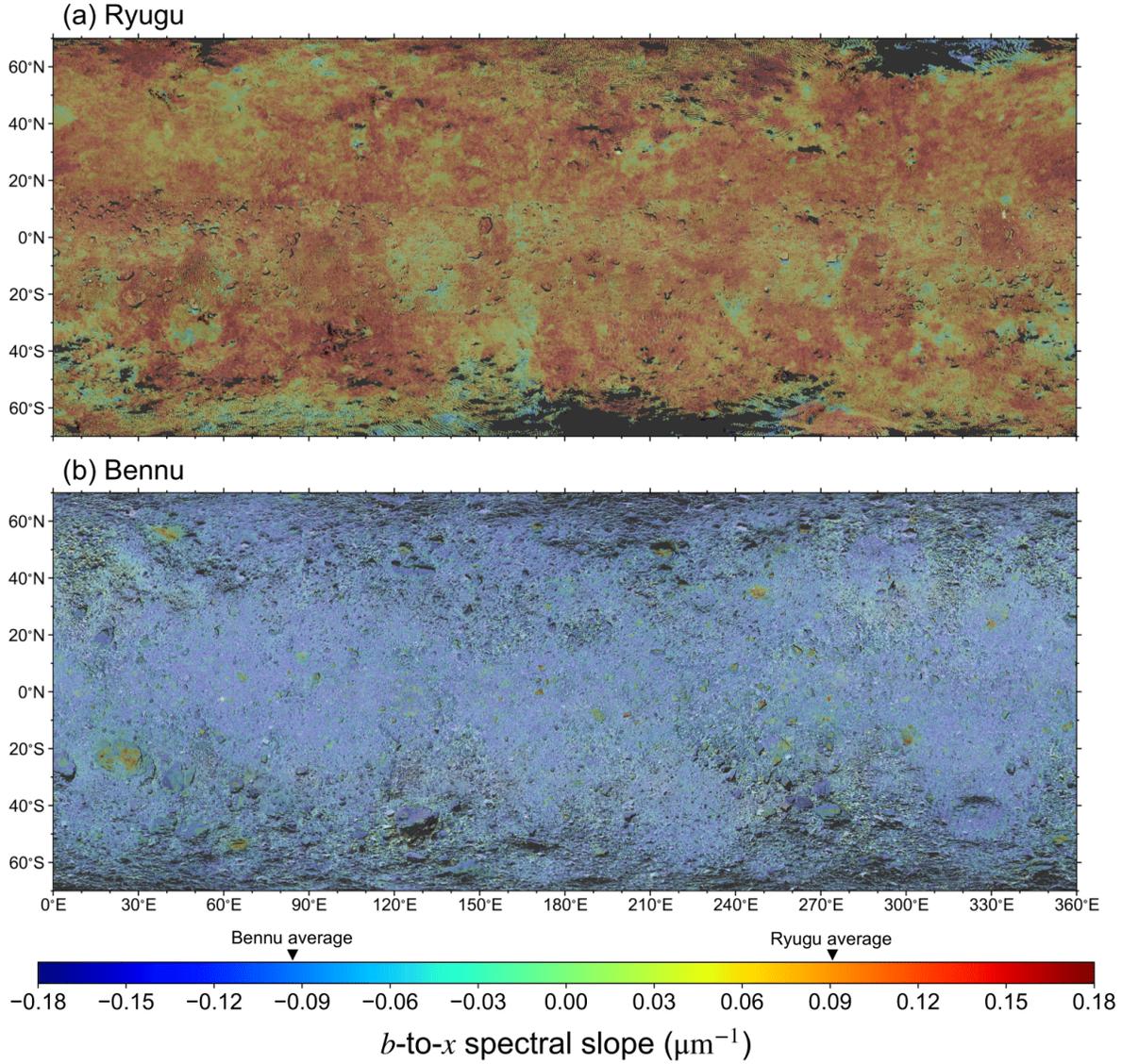

**Figure 2.** Equirectangular projections of *b*-to-*x* band spectral slopes of **(a)** Ryugu and **(b)** Bennu.

### 3.1.1 Global averages

The global average spectra (Fig. 3) show difference between Ryugu and Bennu. The global average *v*-band reflectance ($r'_{norm,v}$) at $(i, e, a) = (10°, 0°, 10°)$ is 0.0287 for Ryugu and 0.0339 for Bennu, indicating that Bennu is brighter than Ryugu by $18.0 \pm 1.5\%$ at *v* band. The uncertainty of 1.5% shows the accuracy of cross calibration between ONC-T and MapCam (Yumoto et al., 2024; hereafter referred to as cross-calibration error); the extent of bias between Ryugu and Bennu data is smaller than this error. Since the spectrum of Ryugu is red sloped and Bennu is blue sloped, the difference in reflectance is smaller at longer wavelength ($11.2 \pm 1.8\%$ at *x* band) and greater at shorter wavelengths ($19.0 \pm 1.6\%$ at *b* band). Nevertheless, Bennu is brighter in all of the *b*, *v*, *w*, and *x* bands. In addition, analysis taking the phase functions of Ryugu and Bennu (Tatsumi et al.,

2020; Golish et al., 2020) into account shows that Bennu is brighter at any phase angles of observation. The global average *b*-to-*x* band spectral slope is 0.0906 μm$^{-1}$ for Ryugu and –0.0933 μm$^{-1}$ for Bennu, indicating that Bennu is bluer than Ryugu by 0.18 ± 0.03 μm$^{-1}$; the uncertainty of 0.03 μm$^{-1}$ shows the cross-calibration error (Yumoto et al., 2024). Thus, we conclude that the average surface material of Bennu is brighter and bluer than that of Ryugu by more than five times the cross-calibration error.

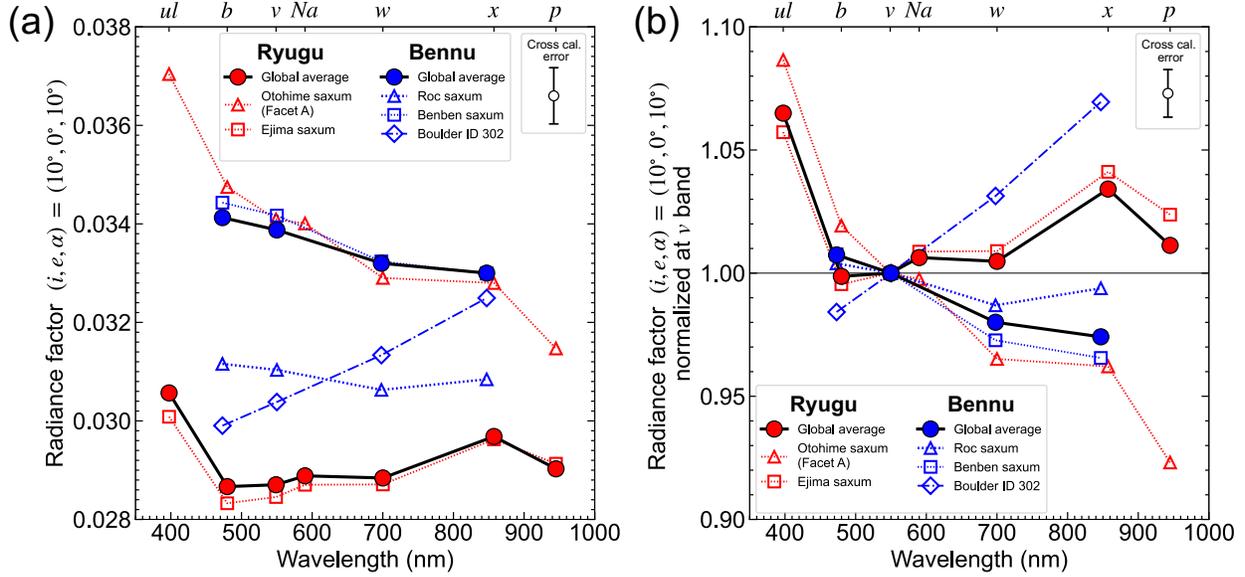

**Figure 3.** Global average spectrum ($r'_{norm,n}$) of Ryugu and Bennu **(a)** without and **(b)** with normalization at *v* band. Dotted lines show the spectra of 1$^{st}$ and 2$^{nd}$ largest boulders: the Otohime saxum (160 m dia.) and Ejima saxum (70 m dia.) on Ryugu and the Roc saxum (150 m dia.) and Benben saxum (70 m dia.) on Bennu. The dash-dotted line labelled "Boulder ID 302" shows the spectrum of the darkest and reddest boulder (10 m dia.) on Bennu (DellaGiustina et al., 2020). The error bar in the upper right corner of each panel indicates the magnitude of cross-calibration error between ONC-T and MapCam, demonstrating that the bias between Ryugu and Bennu data is smaller than this range. The cross-calibration error is ±2% for radiance factor and ±1% for band ratios.

*3.1.2 Latitudinal trends*

The reflectance and spectral slope show gradual changes with latitude, although the global difference between the two asteroids is larger. The equator of Ryugu is brighter than its higher latitudes (Fig. 4a). Additionally, the equator and poles are bluer than the mid-latitudes (Fig. 4b; Sugita et al., 2019; Barucci et al., 2019). These latitudinal differences are statistically significant because they exceed the standard error, as indicated by the shading in Fig. 4. The bright and blue region at the equator creates peaks in Figs. 4a and b with widths of 20º.

The latitudinal trends of Bennu also peak near the equator, indicating that materials near the equator are brighter and bluer than those in the mid-latitudes (Fig.4; DellaGiustina et al., 2020; Barucci et al., 2020). This trend is consistent with observations by the OSIRIS-REx Visible and InfraRed Spectrometer (OVIRS), which showed that the equator is bluer in the 0.44–0.64 μm wavelength range (Li et al., 2021). However, it is worth noting that, OVIRS also found a reversal of this trend (i.e., the equator is redder) in the 1–2 μm range (Li et al., 2021; Clark et al., 2023). The widths and positions of the brighter/bluer peaks near the equator differ compared to Ryugu. The centers of these peaks occur at ~20ºN on Bennu, resulting in asymmetric latitudinal trends between the northern and southern hemispheres. In contrast, the peaks at the equator occur almost perfectly at 0º on Ryugu, and the latitudinal trends are more symmetric. Additionally, the spectral change with latitude occurs more gradually on Bennu, and the peaks near the equator are broader.

The presence of brighter and bluer materials near the equator compared to the midlatitudes on both Ryugu and Bennu can be explained by mass wasting toward or away from the equator. On Ryugu, mass movements likely occur from the equatorial region to the midlatitudes due to the lower geopotential at the equator and the postulated decrease in rotation period (Sugita et al., 2019; Watanabe et al., 2019). Since such mass movements would expose fresher materials at the equator, the brighter/bluer materials in Ryugu's equatorial region should be less weathered, suggesting that space weathering on Ryugu caused darkening and reddening (Sugita et al., 2019; Morota et al., 2020; Tatsumi et al., 2021). In contrast, on Bennu, mass movements likely occurred in the opposite direction (i.e., from the midlatitudes to the equator) due to the higher geopotential at the equator and the increase in rotation period (Nolan et al., 2019; DellaGiustina et al., 2020; Jawin et al., 2020). Thus, the brighter/bluer material at the equator is likely to be more weathered, suggesting that space weathering caused brightening and bluing on Bennu (DellaGiustina et al., 2020; Lauretta et al., 2022).

The more gradual and asymmetric latitudinal trends on Bennu may be due to the more efficient lateral mixing of the surface material. Analysis of surface features indicates that Bennu experienced heterogeneous resurfacing (i.e., mass movements of the surface material not necessarily in the N-S direction) in the recent past (Jawin et al., 2022). Such recent resurfacing likely muted the latitudinal trends formed slowly through asteroid spin-up by mixing the materials laterally. In addition, the older and younger units created by this heterogeneous resurfacing are intertwined on the asteroid's surface, and their distributions are not symmetric between the northern and southern hemispheres (Jawin et al., 2022). This may explain why reflectance and spectral slopes peak at ~20ºN rather than 0ºN on Bennu. The presence of sharp peaks located precisely at the equator on Ryugu (Fig. 4) suggests that such lateral mixing of the surface material might not have been as efficient on Ryugu as it has been on Bennu. The smaller gravity on Bennu may have allowed for easier lateral mixing than on Ryugu.

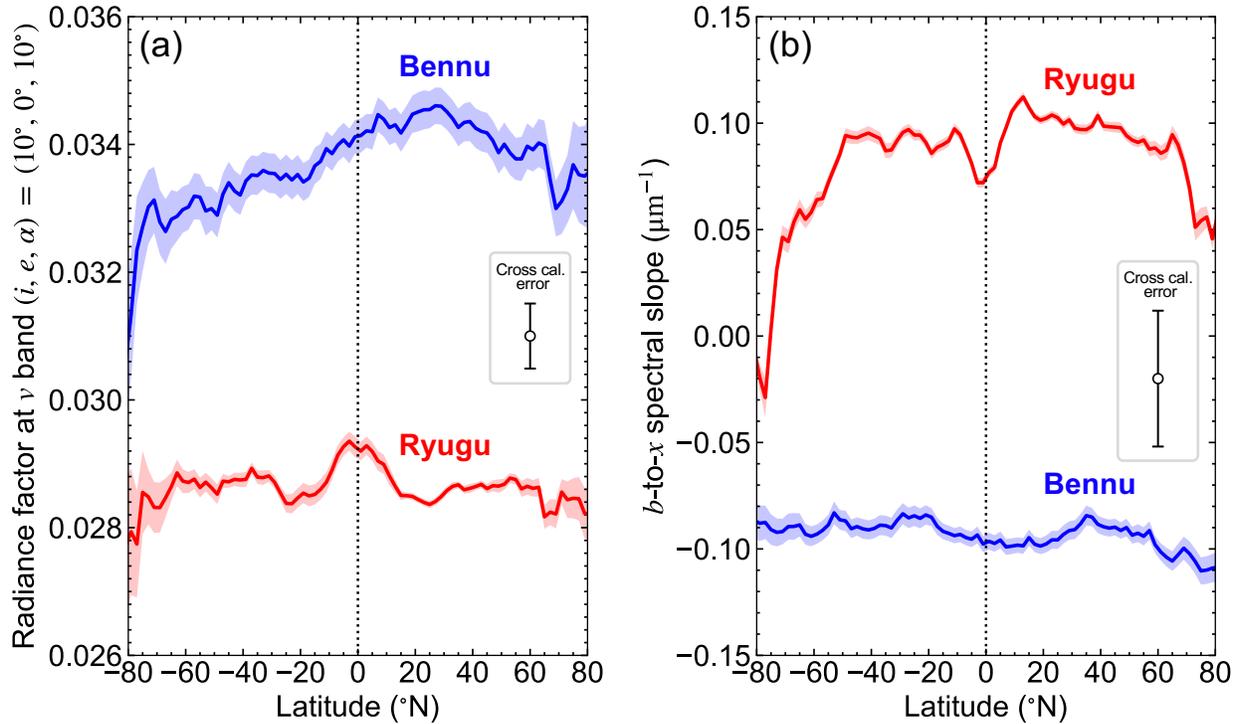

**Figure 4.** Latitudinal variations of **(a)** *v*-band reflectance ($r'_{norm,v}$) and **(b)** *b*-to-*x* spectral slope of Ryugu and Bennu. Solid lines show the average value within each 2º bin, and shaded areas indicate the standard error of the mean. The error bar on the right of each panel indicates the magnitude of cross-calibration error between ONC-T and MapCam, demonstrating that the bias between Ryugu and Bennu data is smaller than this range. The cross-calibration error is ±2% for radiance factor and ±0.03 µm$^{-1}$ for *b*-to-*x* spectral slope.

### *3.1.3 Spatial variability*

Despite the significant difference in global average spectra (section 3.1.1), the spectra of some local features on Ryugu and Bennu show similarities. For instance, the Otohime saxum, the largest boulder on Ryugu, has a spectrum as bright and blue as Bennu's global average (Fig. 3). Although the similarity in their spectral shapes has been previously discussed (Tatsumi et al., 2021), our cross-calibrated spectra allowed us to confirm that their absolute reflectances are also similar. Additionally, the darkest and reddest boulder on Bennu is even redder than Ryugu's average, although its reflectance is not as dark (Fig. 3). Thus, the range of spectral slope variation among meter-scale boulders encompasses the asteroid-to-asteroid difference.

To compare the spectra of the two asteroids on such a local scale, we present histograms of reflectance and spectral slope at 0.3 m spatial resolution (Fig. 5). These histograms show that the spectral distributions of Ryugu and Bennu overlap over a range wider than the magnitude of cross-calibration error. Such a broad overlap shows that certain materials larger than at least 0.3 m have similar spectra between Ryugu and Bennu.

Both asteroids show correlations between reflectance and spectral slope from dark/red to bright/blue as shown by the oblong Gaussian shapes of the 68% contours (Fig. 5c). The different orientations of these Gaussians between Ryugu and Bennu show that most of their spectral variations occur along different spectral trends. The 95% contour curve of Bennu also has a Gaussian-like shape, but interestingly, it has a protrusion toward the distribution of Ryugu; the apex of this protrusion is very close to the global average of Ryugu. This minor population appears to connect the spectral distributions of the two asteroids. In section 3.2.2, we further discuss that the spectra of craters on Bennu lie along this intriguing trend, suggesting that space weathering effects may account for this minor trend.

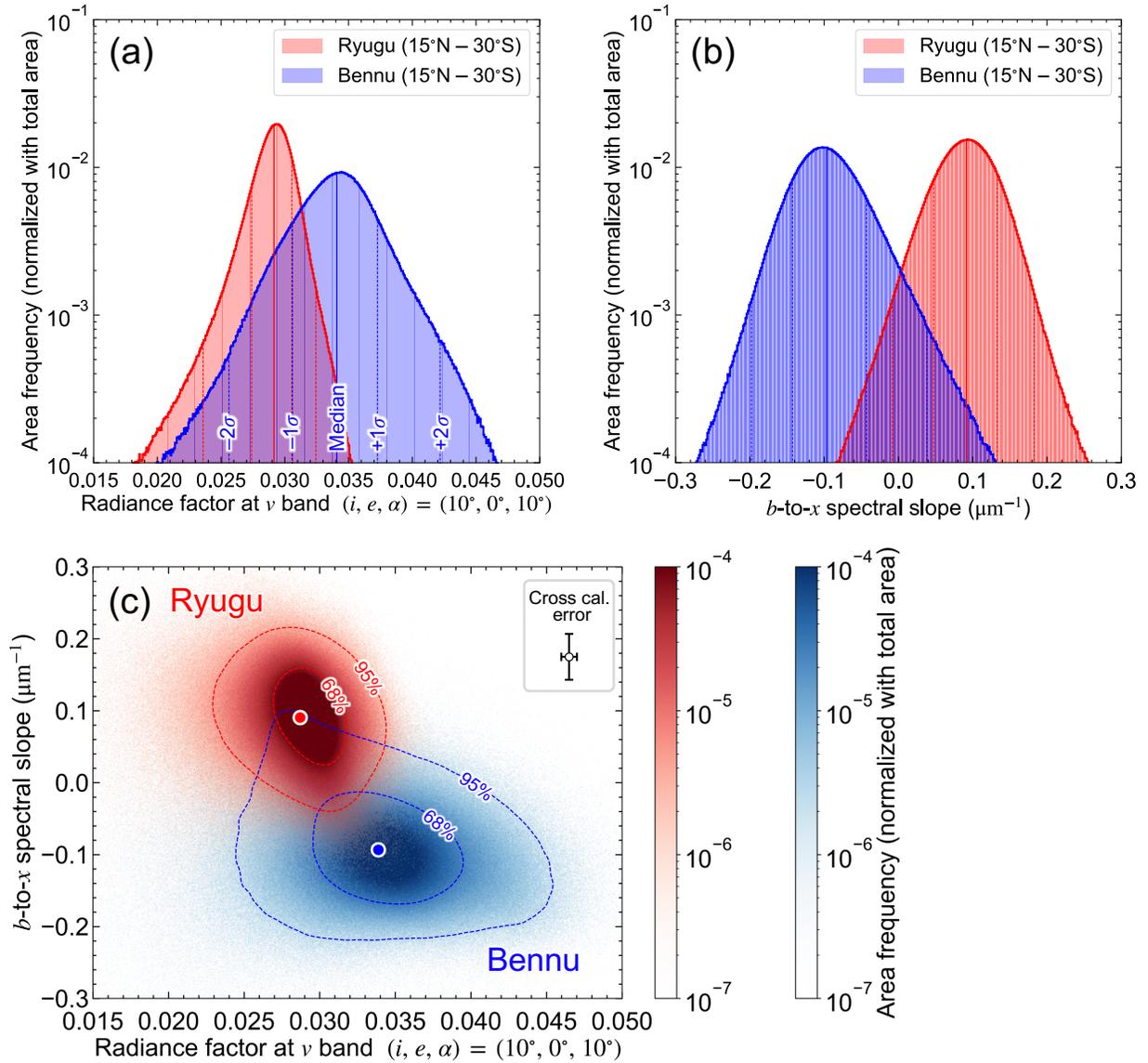

**Figure 5.** Areal frequency histograms of **(a)** $v$-band reflectance ($r'_{norm,v}$) and **(b)** $b$-to-$x$ spectral slope in the 15°N–30°S region of Ryugu and Bennu at 0.3 m spatial resolution. The solid vertical lines show the median, and the dashed vertical lines show the ±1$\sigma$ and ±2$\sigma$ ranges. **(c)** A contour map showing the correlation between the reflectance and spectral slope. The dashed curves show 68 and 95% percentiles. Points at the center of each distribution show the global (90°N–90°S) averages. The error bar in the upper right corner indicates the magnitude of cross-calibration error between ONC-T and MapCam, demonstrating that the bias between Ryugu and Bennu data is smaller than this range. The cross-calibration error is ±2% for radiance factor and ±0.03 μm$^{-1}$ for $b$-to-$x$ spectral slope.

We next compare the spectral heterogeneity on Ryugu and Bennu as functions of spatial scales. DellaGiustina et al. (2020) showed that the degree of spectral heterogeneity is larger when observed at 2 m/pix than at 64 m/pix for both Ryugu and Bennu. Indeed, the spatially binned equirectangular projections (Fig. 6) show that heterogeneities are larger at 1 m/pix than at 20 m/pix. By comparing the heterogeneity between Ryugu and Bennu, we observe that the spectral slope exhibits greater variability on Ryugu than on Bennu at 20 m/pix, while the heterogeneity on Bennu seems to become comparable to that on Ryugu at 1 m/pix (Fig. 6b). This observation implies that the spatial scales of materials producing the spectral variation may be different between Ryugu and Bennu.

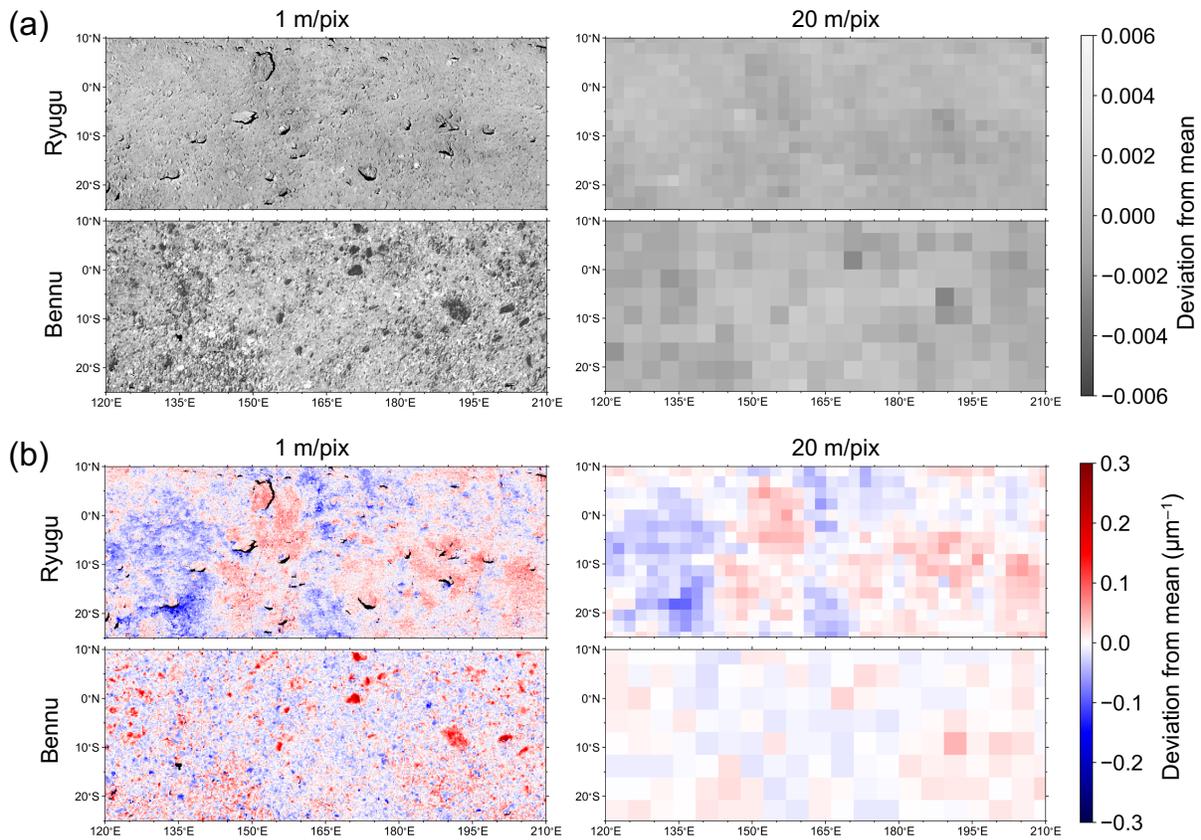

**Figure 6. (a)** Reflectance at $v$ band ($r'_{norm,v}$) and **(b)** $b$-to-$x$ spectral slope maps of Ryugu and Bennu shown in deviations from their mean values. These maps show the spectral heterogeneity at spatial scales of **(left)** 1 m/pix and **(right)** 20 m/pix.

We quantitatively investigated the spectral variations at different spatial scales by calculating the standard deviations of reflectance and spectral slope while gradually changing the spatial scale of binning (Fig. 7). The variation in reflectance and spectral slope show a power-law increase as the spatial scale becomes smaller. The fitted power-law indices of the reflectance variation (Fig. 7a) are –0.372 ± 0.005 for Ryugu and –0.322 ± 0.002 for Bennu in the 1–10 m/pix range. The similarity in the power-law indices suggests that the reflectance variation on each asteroid's surface is primarily caused by surface features with similar size distribution. For instance, it may be caused by the reflectance variation among boulders because their size distributions have similar power-law indices: –2.65 ± 0.05 for Ryugu (Michikami et al., 2019) and –2.9 ± 0.3 for Bennu (DellaGiustina et al., 2019). The higher abundance of bright boulders on Bennu (DellaGiustina et al., 2020) may explain the larger reflectance variation on Bennu than on Ryugu at all spatial scales. We observe an apparent transition in power-law indices around 10 m/pix on both asteroids; the power-law indices in the 10–50 m/pix range are –0.167 ± 0.002 for Ryugu and –0.229 ± 0.004 for Bennu. This transition at ~10 m/pix may be due to the change in the dominant surface features from individual boulders to large geologic features (e.g., ridges, troughs, and craters).

In contrast, the power-law indices of the spectral slope variation (Fig. 7b) are significantly different: –0.145 ± 0.003 for Ryugu and –0.317 ± 0.003 for Bennu (1–10 m/pix range). As a result, the magnitude of heterogeneity reverses at ~1 m scale. The heterogeneity is larger for Bennu at scales smaller than 1 m, but Ryugu is more heterogeneous at scales larger than 1 m. The greater homogeneity of Bennu at large spatial scales suggests a higher mixing efficiency of surface materials over long distances, consistent with the findings in section 3.1.2.

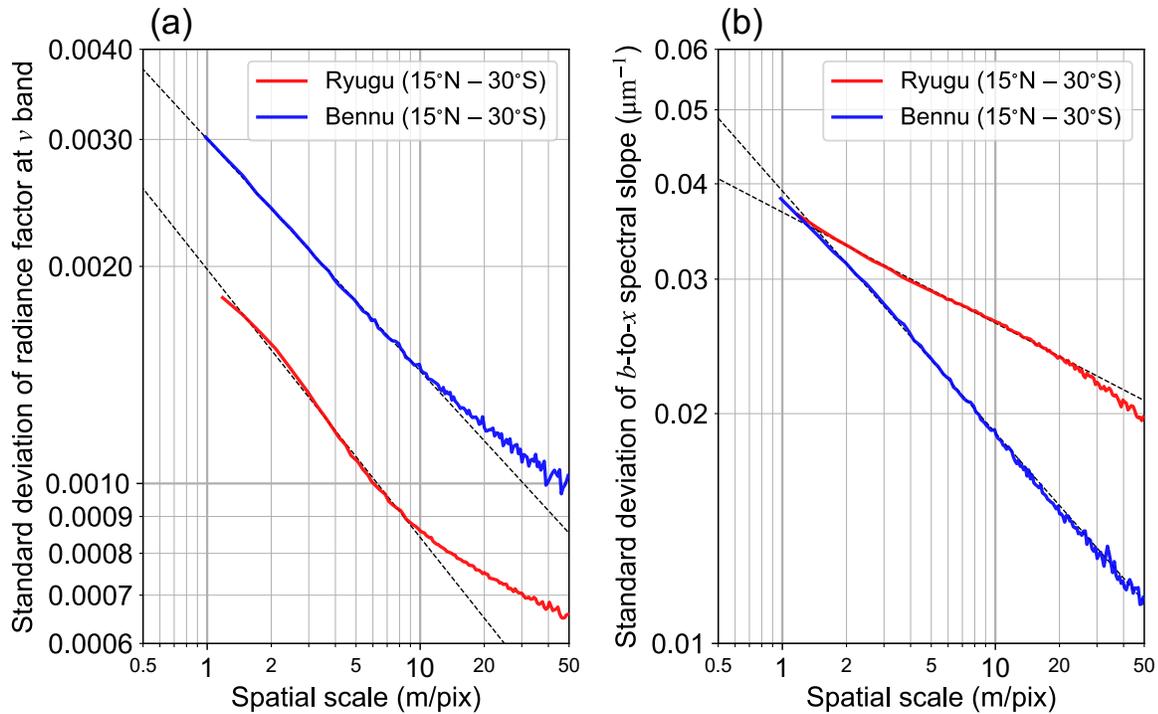

**Figure 7.** Standard deviations of **(a)** $v$-band reflectance ($r'_{norm,v}$) and **(b)** $b$-to-$x$ spectral slope within 15ºN to 30ºS as a function of spatial scale. The dashed lines show the power laws fitted

within the range of 1–10 m/pix.

## 3.2 Comparison of crater spectra

### 3.2.1 Similarity in the spectra of freshest craters on Ryugu and Bennu

We compare the spectral distributions of craters before and after cross calibration in Fig. 8. An offset observed between the two apparently parallel spectral distributions of craters on Ryugu and Bennu in the pre-cross-calibrated data (Fig. 8a) diminished after cross calibration (Fig. 8b). This indicates that the systematic error between the two imagers caused the offset in the pre-cross-calibrated data. With the improved precision achieved through cross calibration, we conclude that the spectral distributions of craters on Ryugu and Bennu are adjacent to each other without a significant gap.

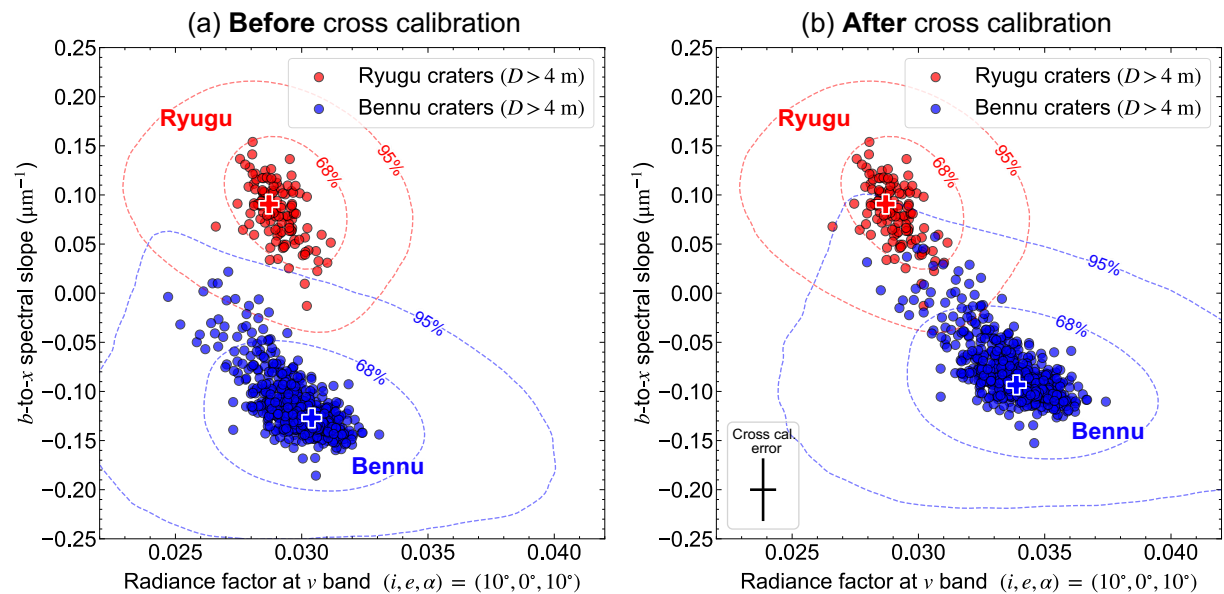

**Figure 8.** Reflectance–spectral slope distributions of craters **(a)** before and **(b)** after cross calibration. Symbols "+" show the global average, and the dashed lines show the 68 and 95% percentiles within the 15ºN to 30ºS region observed at 0.3 m/pix (same as Fig. 5c). The error bar in the lower left corner of panel (b) indicates the magnitude of cross-calibration error between ONC-T and MapCam, demonstrating that the bias between Ryugu and Bennu data is smaller than this range. The cross-calibration error is ±2% for radiance factor and ±0.03 $\mu m^{-1}$ for $b$-to-$x$ spectral slope.

The spectra of craters on Ryugu and Bennu become closer to each other at smaller crater sizes. The spectra of craters larger than 120 m are distributed close to the global average for both asteroids (Fig. 9a). At smaller size ranges, the population of craters with spectra brighter and bluer than the global average increases for Ryugu, while those with darker and redder spectra increase for Bennu (Fig. 9a). This trend becomes more evident after averaging the spectra across different size ranges (Fig. S3). Fig. S3 shows that the average crater spectra gradually become brighter and bluer for Ryugu and darker and redder for Bennu as their size decreases, with the only exception being the reversal of the trend below 10 m in size for the spectral slopes of Ryugu craters. Thus, the spectra of smaller craters on Ryugu and Bennu generally become closer together. Consequently, the spectral distributions of craters on Ryugu and Bennu begin to converge at a size range of 4–15 m (Fig. 9a). The $b/v$ band ratio and $v$-to-$x$ spectral slope of craters also become closer at smaller crater sizes (Fig. S2).

Such a correlation between crater spectra and size suggests a darkening–reddening space weathering trend for Ryugu and a brightening–bluing trend for Bennu, both of which are consistent with previous reports (Sugita et al., 2019; Morota et al., 2020; DellaGiustina et al., 2020; Lauretta et al., 2022). The reversal in the trend seen in the spectral slopes of Ryugu craters smaller than 10 m might indicate that the space weathering trend on Ryugu is not unidirectional; for example, bluing occurs over short timescales, while reddening becomes prominent over longer timescales. However, it is also possible that this reversal reflects certain stratigraphy within the shallow (<1 m depth) subsurface rather than space weathering. Morota et al. (2020) also observed that the smallest craters on Ryugu are not the bluest. They suggested that this may be due to the presence of a meter-thick redder layer in the uppermost surface; craters smaller than 10 m are unlikely to expose the blue materials deeper than this layer, considering the typical crater depth-to-diameter ratio of 0.1 (Noguchi et al., 2021). Such a reversal of the trend may explain why the tail of the crater spectral distribution (Fig. 9a) is shorter for Ryugu than for Bennu.

The large spectral variation among the smallest (4–15 m) craters investigated in this study (Figs. 9a and S3) likely reflects their varying degrees of space weathering. This is because the retention ages of these smallest craters are still comparable to or longer than the typical timescales of space weathering. For instance, the retention ages of 10 m-sized craters are as long as $10^3$–$10^6$ years (Takaki et al., 2022). Noble gas measurements of returned samples also indicate that Ryugu samples remained in the upper 1 m subsurface for $5 \times 10^6$ years (Okazaki et al., 2020), providing additional evidence for the long retention age of 10 m-sized (i.e., 1 m-depth) craters. In contrast, space weathering can occur on shorter or comparable timescales. For instance, solar wind irradiation causes weathering over timescales of $10^3$–$10^4$ years (Lantz et al., 2017) and the removal of fine grains through electrostatic levitation typically occurs over $10^4$–$10^5$ years (Hsu et al., 2022). These timescales suggest that not all of the smallest craters investigated in this study are fresh; some could have been exposed on the surface long enough to undergo weathering processes.

Hence, the darkest/reddest craters on Ryugu and the brightest/bluest craters on Bennu, located at the tail ends of the spectral distributions in Fig. 9a, are likely the freshest ones. These freshest craters on Ryugu and Bennu overlap in Fig. 9a, indicating that their reflectance and spectral slope are indistinguishable. Fig. 10 additionally shows that the spectra of these freshest craters are similar across the entire $b$-to-$x$ band range. The freshest craters on Ryugu appear to have a larger absorption feature at the $w$ band than those on Bennu, but the difference is within the cross-calibration error. Therefore, we conclude that the spectra of the freshest craters on Ryugu and Bennu are indistinguishable in the $b$, $v$, $w$, and $x$ bands, at least within the accuracy achieved in our cross calibration (Yumoto et al., 2024). The Hokioi crater on Bennu, which contains the Nightingale sample collection site, also has a spectrum indistinguishable from that of the freshest craters on Ryugu (Fig. 10).

Our findings, indicating that the spectra of craters on both asteroids become closer as size decreases and that the spectra of the freshest craters are indistinguishable, suggest that Ryugu and Bennu evolved from materials with similar visible spectra (Fig. 9b). The similarity in the initial spectra leads us to further assume that the starting materials of Ryugu and Bennu may have been similar. However, this is not conclusive, as compositionally different materials can exhibit similar visible spectra.

We note that while the trend observed among natural craters suggests fresher materials to be brighter on Ryugu, the floor and ejecta of the artificial crater created by the Small Carry-on Impactor (SCI) appear darker than the surrounding area (Arakawa et al., 2020). This discrepancy may be attributed to the transient presence of fine-grained material within the SCI crater. The observed phase function of the SCI crater have suggested that its dark reflectance is caused by the presence of fine dust (Honda et al., 2022). Particle size distributions within the floor and ejecta curtain of the SCI crater also indicate that it consists of finer particles compared to the average surface (Kadono et al., 2020; Wada et al., 2021). Darkening by such fine dusts is likely to be a transient process because dark ejecta, such as those found around the SCI crater, are absent around even the freshest natural craters investigated in this study (Honda et al., 2022).

We also note that the trends observed in this study are partly inconsistent with those obtained from OVIRS data (Deshapriya et al., 2021), but this inconsistency likely originates from the fact that the OVIRS spectra of certain small craters were observed with different geometries. Our result shows that smaller craters on Bennu are *darker* at $v$ band nm, whereas the analysis of OVIRS data by Deshapriya et al. (2021) reported that they are *brighter* at 550 nm. The crater reflectance reported in these two studies are highly correlated for craters observed under the same 8° phase angle condition; correlation coefficient is >0.7. However, the reflectance of certain small craters reported by Deshapriya et al. (2021), observed at high phase angles of ≥30° during the reconnaissance flybys, deviate from this consistency and appear to be systematically brighter (Fig.

S4). Although the exact cause of this discrepancy remains unidentified, our results suggest that the trend reported in Deshapriya et al. (2021) may involve inaccuracies due to the biased observation geometry of small craters.

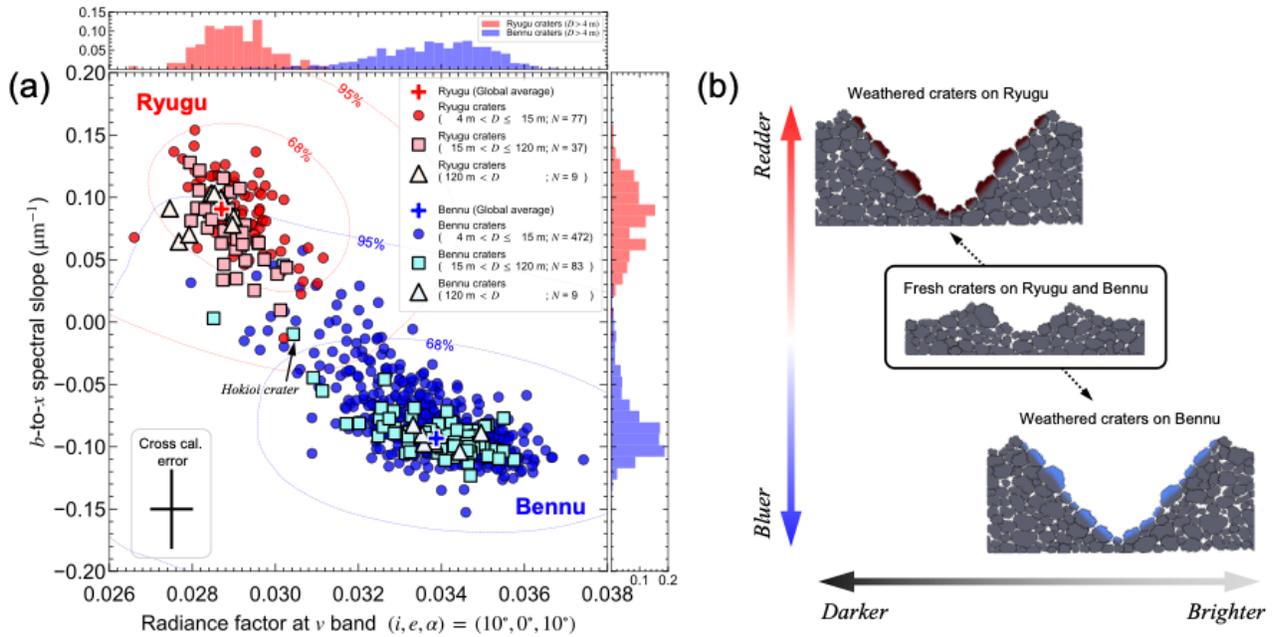

**Figure 9.** Reflectance–spectral slope distribution of craters with different size ranges on Ryugu and Bennu. **(a)** Spectral distribution of craters. The histograms on the top and right show the frequency of craters normalized with total count. Symbols "+" show the global average, and the dashed lines show the 68 and 95% ($1\sigma$ and $2\sigma$) percentiles within the 15ºN to 30ºS region observed at 0.3 m/pix (same as Fig. 5c). The error bar in the lower left corner indicates the magnitude of cross-calibration error between ONC-T and MapCam, demonstrating that the bias between Ryugu and Bennu data is smaller than this range. The cross-calibration error is ±2% for radiance factor and ±0.03 μm$^{-1}$ for *b*-to-*x* spectral slope. **(b)** Schematic illustrations showing the implications of **(a)**. The likely freshest materials of Ryugu and Bennu found inside certain small craters exhibit similar reflectance and spectral slope (see section 3.2.1). However, space weathering changed the spectra of Ryugu and Bennu in the completely opposite direction along the same straight trend line shown in dashed arrows (see section 3.2.2). Consequently, weathered materials found in large craters are darker and redder on Ryugu while those of Bennu are brighter and bluer.

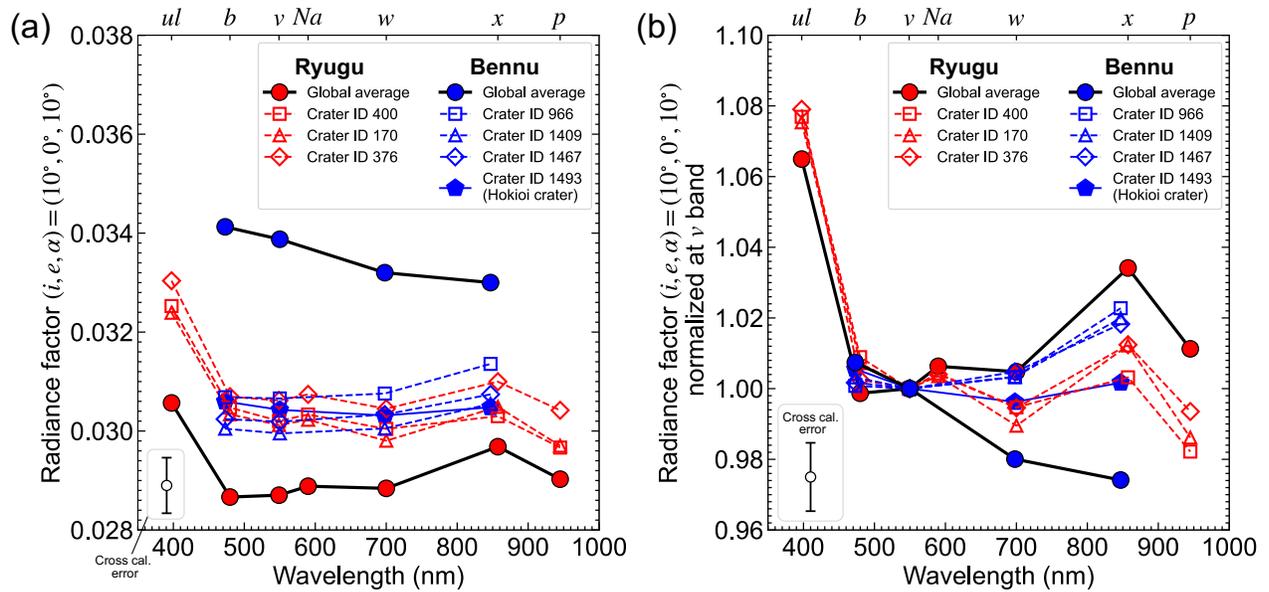

**Figure 10.** Spectra of the three reddest (i.e., largest *b*-to-*x* spectral slope) craters on Bennu and three bluest (i.e., smallest *b*-to-*x* spectral slope) craters on Ryugu **(a)** without and **(b)** with normalization at *v* band (550 nm). We also show the spectra of the Hokioi crater: the crater containing the Nightingale sample collection site on Bennu. The error bar in the lower left corner of each panel indicates the magnitude of cross-calibration error between ONC-T and MapCam, demonstrating that the bias between Ryugu and Bennu data is smaller than this range. The cross-calibration error is ±2% for radiance factor and ±1% for band ratio.

### 3.2.2 *Similarity in the trend lines of space weathering on Ryugu and Bennu*

We examined whether the space weathering trends of Ryugu and Bennu, as suggested by the observed size–spectra relation of craters (section 3.2.1), follow distinct trend lines or converge onto a common trend line in the reflectance-spectral slope diagram. We obtained the trend lines by calculating the cumulative-average spectra of craters as a function of their diameter, plotting the averaged data in the reflectance–spectral slope diagram, and fitting them with a linear function (Fig. 11).

On Ryugu, we observe a gradual darkening–reddening trend from a crater size of 10 m to 80 m. The fitted slope and intercept of this trend are $-34 \pm 6$ μm$^{-1}$ and $1.1 \pm 0.2$ μm$^{-1}$, respectively. As discussed in section 3.2.1, the reversal of the reddening trend observed for craters smaller than 10 m may reflect the stratigraphy within the shallow subsurface. This trend line of space weathering appears to align with the global spectral distribution, as represented by the 68% contour in Fig. 11, suggesting that most of the spectral variation on Ryugu reflects the varying degrees of space weathering effects.

On Bennu, we observe a gradual brightening–bluing trend from a crater size of 4 m to 80 m. The slope and intercept of the trend are –32 ± 5 µm$^{-1}$ and 1.0 ± 0.2 µm$^{-1}$, respectively. Unlike Ryugu, this trend line of space weathering on Bennu is oriented in a direction different from that of the global spectral distribution, as represented by the 68% contour in Fig.11, but rather aligns with the minor trend discussed in section 3.1.3. The result suggests that most of the spectral variation on Bennu reflects the diversity in starting materials rather than variation in the degree of space weathering.

These results show that the observed trend lines, which are suggestive of space weathering, are consistent between Ryugu and Bennu within their fitting errors in the reflectance–spectral slope diagram. This result suggests that the spectra of Ryugu and Bennu evolved in *completely* opposite directions along the same straight trend line, without any bend in the middle (Fig. 9b).

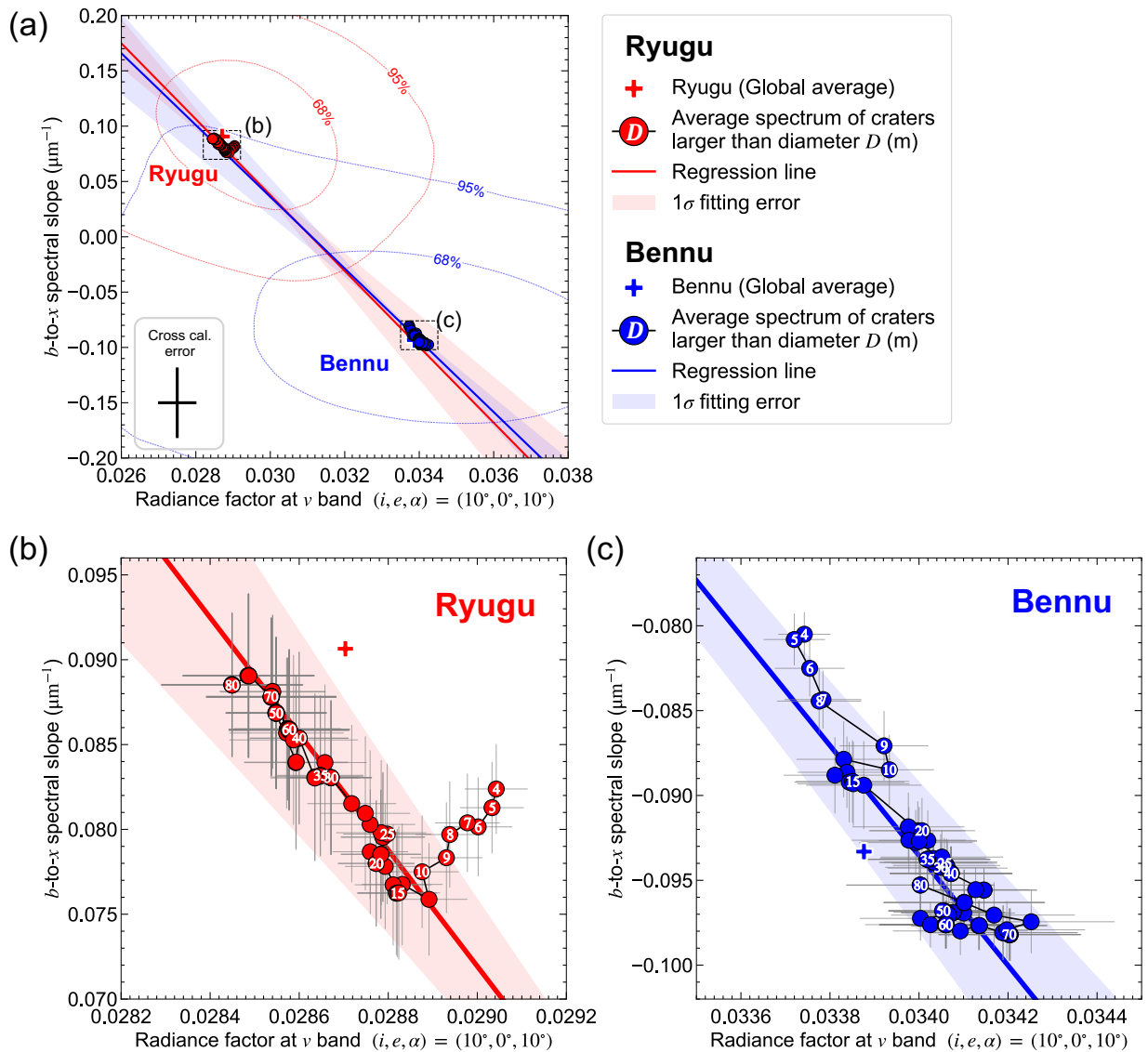

**Figure 11**. Trend lines of space weathering on Ryugu and Bennu in the reflectance–spectral slope diagram calculated from the gradual change of crater spectra with size. The plots **(b)** and **(c)** magnify the dashed boxes shown in **(a)**. Each point represents the average spectra of craters with size larger than $D$ (m). The white numerals associated with each marker indicate the thresholds diameter $D$. We vary the threshold diameter from 80 m to 4 m with a step size of 1 m. The thin error bar for each plot shows the standard error. The solid lines show the result of linear fitting and the hatch show the $1\sigma$ fitting error. We fitted the data in the crater size range of 10–80 m for Ryugu and 4–80 m for Bennu. Symbols "+" show the global average, and the dashed lines show the 68 and 95% ($1\sigma$ and $2\sigma$) percentiles within the 15ºN to 30ºS region observed at 0.3 m/pix (same as Fig. 5c). The thick black error bar in the lower left corner of panel (a) indicates the magnitude of cross-calibration error between ONC-T and MapCam, demonstrating that the bias between Ryugu and Bennu data is smaller than this range. The cross-calibration error is ±2% for radiance factor and ±0.03 µm$^{-1}$ for *b*-to-*x* spectral slope.

## 4. Implications for space weathering processes on Ryugu and Bennu

We discuss the possible causes for the opposite space weathering trends on Ryugu and Bennu based on the new findings obtained from the analyses of crater spectra. In section 4.1, we first discuss whether the opposite trends can be accounted for by chemical/mineralogical change due to solar wind/micrometeorite irradiations. Alternatively, in section 4.2, we explore the possibility that these trends may also be explained by changes in the physical properties of the asteroid surface. Although our results are consistent with both of these space weathering processes, the fact that space weathering trends on Ryugu and Bennu occur along a common trend line without a bend (section 3.2.2) may be more readily explained by the latter.

### 4.1 Space weathering involving chemical/mineralogic change

Laboratory simulations of space weathering, involving ion and pulsed laser irradiation, suggest that space weathering likely induces certain chemical and mineralogical changes. Products formed after such processes include nano to micron-sized Fe/Fe-oxides/Fe-sulfides, dehydrated phyllosilicates (Noguchi et al., 2023), and carbonized organics.

The properties (e.g., chemical composition, grain size, amorphization) of these space weathering products likely determine the spectral change. For instance, spectral reddening may occur on silicate compounds due to the production of npFe$^0$ (Sasaki et al., 2001; Hapke, 2001), while bluing may occur on carbon-rich materials due to the carbonization of organics (Moroz et al., 2004; Hendrix & Vilas, 2019) and the production of micron-sized carbon particles (µpC; Kaluna et al., 2017). Although the same weathering product can induce different spectral effects when the initial spectra differ (e.g., productions of npFe$^0$ induce darkening on bright, semi-transparent materials but its effect on the spectra of dark, optically opaque materials may not be as significant; Pieters, 2000; Rivkin et al., 2002), this is not likely the case for Ryugu and Bennu, as their initial spectra are similar (section 3.2.1). Thus, our results suggest that certain properties of space weathering products differ between Ryugu and Bennu.

The space weathering products contributing to the spectral change may differ between Ryugu and Bennu due to their possible differences in (1) composition or (2) weathering agent/dose, as discussed in the following.

(1) Laboratory experiments by Lantz et al. (2017) show that spectral changes by space weathering depend on the composition of the material. Darkening/reddening by npFe$^0$ dominates the spectral changes of chondrule-rich carbonaceous chondrites (e.g., CO, CV), while brightening/bluing by the carbonization of organics may dominate that of matrix-rich carbonaceous chondrites (e.g., CM, CI). However, it is still uncertain whether the composition of Bennu materials is similar to the CI-like Ryugu material (Yokoyama et al., 2022). The presence of magnetite inferred from the thermal infrared spectra on Bennu (Hamilton et al., 2019) supports their similarity to CI chondrites. In contrast, the wide 2.7 μm absorption feature with a hockey-stick shape on Bennu suggests a closer match with CM chondrites (Hamilton et al., 2019). Additionally, the thermal infrared spectrum of Bennu best matches that of a CR1 chondrite, GRO 95577 (Hamilton et al., 2022).

(2) Laboratory experiments also show that asteroids with similar compositions may undergo different spectral effects of space weathering when the agents or doses of space weathering differ. Thus, we cannot rule out the possibility that Ryugu and Bennu evolved from materials with similar composition. For instance, researchers have shown that ion irradiation (Lantz et al., 2015), pulsed laser irradiation (Matsuoka et al., 2015), and heating (Hiroi et al., 1996) experiments on a Murchison meteorite all result in different spectral effects. This variability is important to consider because the darkening and reddening observed on Ryugu, which has a CI-like composition (Yokoyama et al., 2020), has not been reproduced by *ion* irradiation on CI chondrite pellets (Lantz et al., 2017), but has recently been reproduced by *laser* irradiation on CI simulant powders (Prince & Loeffler, 2022). Changing the species (Vernazza et al., 2013) or the energy (Nakamura et al., 2020) of irradiated ions can also lead to opposite spectral effects. Additionally, the spectral effect of space weathering is not necessarily linear with respect to the total dose (i.e., the total number of irradiated ions or laser pulses). In fact, increasing the irradiation dose can sometimes even invert the spectral change. This has been demonstrated by helium (He$^+$) exposure experiments on Murchison chips (Nakamura et al., 2020) and pulsed laser irradiation on carbonaceous chondrite simulants (Kaluna et al., 2017; Prince & Loeffler, 2022). The agent and dose of space weathering may vary among asteroids depending on their orbital evolutions. While there is no strong evidence to suggest significant differences in the orbital evolutions of Ryugu and Bennu, it has been proposed that Ryugu may have undergone an orbital excursion near the Sun (Morota et al., 2020). We should also note that differences in timescales of resurfacing processes may contribute to the color–age trends observed on asteroid surfaces. However, this is unlikely for Ryugu and Bennu, as their surface ages estimated from crater size frequencies are similar (Takaki et al., 2022; Bierhaus et al., 2022).

The mineralogy and elemental abundance of the returned samples would directly reveal the compositional similarities and differences between Ryugu and Bennu materials. Such analysis should be able to determine which of the two hypotheses (i.e., differences in composition or weathering agent/dose) or other factors caused the opposite space weathering trends. Thus, our remote sensing data underscore the value of geochemical comparison between Ryugu and Bennu samples, as the results could significantly alter our interpretation of C-complex asteroid spectra (see also section 5).

Another key issue that needs to be addressed in future studies is identifying whether a process exists that can explain the *completely* opposite space weathering trends observed on Ryugu and Bennu. We showed in section 3.2 that the space weathering trends of Ryugu and Bennu occur in opposite directions but along a common straight trend without a bend. This is a strong observational constraint because the formation of different space weathering products would likely alter the spectra along different trend lines, making it unlikely for these trends to coincidentally align. Reproducing the trend line shared by Ryugu and Bennu will be important in future simulations of space weathering in laboratories.

## 4.2 Space weathering involving change in physical properties

The evolutions of physical properties (e.g., thickness of dust coating and grain size) of the surface material is another viable space weathering process that may explain the trends observed on Ryugu and Bennu.

Laboratory experiments using carbonaceous chondrites show that differences in the thickness of dust coating or grain size greatly affect the optical spectrum (e.g., Johnson & Fanale, 1973; Cloutis et al., 2011a; 2011b; Kiddell et al., 2018). Fig. 12 compares these effects with the spectral variations among Ryugu and Bennu craters. Although the trends of spectral change depend on the meteorite type, Fig. 12 shows that variations in the thickness of dust coating or grain size on scales of just 10–100 μm typically account for the range of spectral variations observed among craters on Ryugu and Bennu.

Although the diurnal temperature curves of Ryugu and Bennu show that fine dust particulates are mostly absent on both asteroid surfaces (Grott et al., 2019; Biele et al., 2019;

Rozitis et al., 2020), the presence of thin dust layers with a thickness of a few tens of microns is not inconsistent with the observations. Simulations of the temperature curves using two-layer models and lateral mixing models show that the presence of a dust layer with <50 μm thickness or dust coverage by <5–10% area cannot be ruled out (Rozitis et al., 2020). Furthermore, the emissivity drop at 6 μm in the mid-infrared spectra suggests the thin accumulations of fine particles (Hamilton et al., 2021; Hamm et al., 2022).

The grain size of asteroid surfaces may become finer over time due to the disruption of boulders by meteoroid impacts and thermal fragmentation, while finer particulates may be lost due to levitation by electrostatic forces and ejection by high-speed impacts. The balance between these processes determines how the grain size distribution evolves on the asteroid's surface. Asteroid size is one of the important parameters for this balance. Model calculations show that grains larger than millimeter size cover more than half of a 1 km-sized asteroid surface, while the surface of a 10-km-sized asteroid is dominated by fine micron-sized particulates (Hsu et al., 2022). Similarly, the evolution of grain size may occur differently between the 0.5-km-sized Bennu and the 1-km-sized Ryugu and may explain their opposite space weathering trends. For instance, the abundance of fine particulates causing darkening and reddening (Johnson & Fanale, 1973; Morota et al., 2020) may have increased on Ryugu, while they may have been preferentially lost on Bennu, causing its brightening and bluing. The particle ejection events observed on Bennu (Lauretta et al., 2019b) may have also driven the loss of fine particulates. In addition, the possible difference in the mechanical strength of boulders may have contributed to the less efficient productions of fines on Bennu; the estimates for mechanical strength range from 0.44 to 1.70 MPa on Bennu (Ballouz et al., 2020), while those for Ryugu range from 0.2 to 0.28 MPa (Grott et al., 2019).

The common trend line of space weathering observed between Ryugu and Bennu (section 3.2.2) may be more consistent with such physical effects than with chemical/mineralogical changes. This is because a continuous change in a single physical quantity, such as an increase or decrease in grain size, would likely alter the spectra along the same continuous trend line. One caveat to consider is that such grain size effect may not account for the different 2.7-μm features of Ryugu and Bennu (section 4.1). Laboratory experiments suggest that variations in grain size do not cause wide CM-type absorption bands to change into sharp CI types, and vice versa (Cantillo et al., 2023). We also note that the variation in meteorite spectra observed across different grain size ranges may reflect compositional fractionation (Cloutis et al., 2011b). This is because different materials within a meteorite have varying resistance to crushing, resulting in the separation of these materials at different grain sizes.

We can test if such space weathering mechanism can explain the observations by measuring the spectra of returned samples with different grain sizes and observing if their spectra align with the space weathering trends observed by remote sensing.

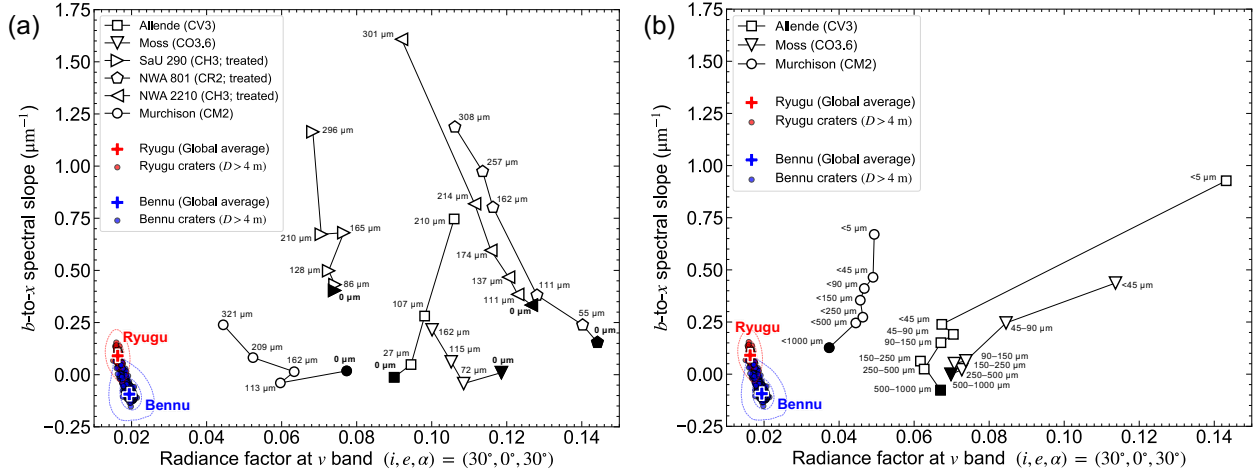

**Figure 12**. Reflectance–spectral slope distributions of craters on Ryugu and Bennu compared with spectral change by **(a)** thickening of fine dust powder coating on meteorite slabs and **(b)** decreasing size of meteorite grains; the meteorite data are from Kiddell et al. (2018). The labels associated with each meteorite spectrum show the thickness of powder coating in (a) and the range of grain size in (b). The grain sizes of fine dusts in (a) are <5 μm for Allende and Murchison and <45 μm for others. Meteorites labelled with "treated" in the legends indicate that rust formed by terrestrial alteration has been removed using thiol solutions. Symbols "+" show the global average of Ryugu and Bennu, and the dashed lines show the 68 and 95% (1$\sigma$ and 2$\sigma$) percentiles within the 15ºN to 30ºS region observed at 0.3 m/pix (same as Fig. 5c).

## 5. Implications for the spectral variation among C-complex asteroids

We discussed in section 4 that the processes accounting for the space weathering trends on Ryugu and Bennu still have many possibilities and need to be tested by analysis of returned samples. Nevertheless, the results of this study imply a clear implication for the C-complex asteroids: the heterogeneity of spectral slopes among C-complex asteroids may have increased after space weathering.

Since our results suggest that Cb-type Ryugu and B-type Bennu had similar initial spectra (section 3.2.1), these two asteroids demonstrate that space weathering is likely the major cause of their different disk-averaged spectra. To put other C-complex asteroids in the context, we compared the range of spectral variation among craters on Ryugu and Bennu with that among the entire C-complex asteroid population (Figs. 13 and S2). We calculated the spectral slopes and band ratios of the C-complex asteroids using the spectra observed by the Eight Color Asteroid

Survey (ECAS; Zellner et al., 1985; Zellner et al., 2020) and the Small Main-Belt Asteroid Spectroscopic Survey, Phase II (SMASS II; Bus & Binzel, 2002a; Bus & Binzel, 2020). We took the geometric albedos from the Wide-field Infrared Survey Explorer (WISE) data (Wright et al., 2010). We compared the geometric albedo of C-complex asteroids with the normal albedo (i.e., radiance factor at $(i, e, \alpha) = (0°,0°,0°)$) of craters on Ryugu and Bennu calculated using equation 2; we can compare these two because $i$ and $e$ have negligibly small effects near $\alpha=0°$ (Golish et al., 2020). The analyzed C-complex asteroids were larger than 4 km in size, with a median size of 33 km.

Figs. 13 and S2 show that the range of spectral slope change on Ryugu and Bennu due to space weathering effects is nearly as large as the spectral slope variation across the C-complex asteroid population. Theis result suggests that space weathering likely expanded the spectral variation of C-complex asteroids. This further implies that C-complex asteroids may have formed from materials with more uniform spectral slopes than the variation observed today by ground-based telescopes. In contrast, the range of reflectance change observed on Ryugu and Bennu is smaller than the albedo variation among the C-complex asteroids. This suggests that most of the albedo variation in C-complex asteroids reflects their diversity in starting materials rather than their space weathering history.

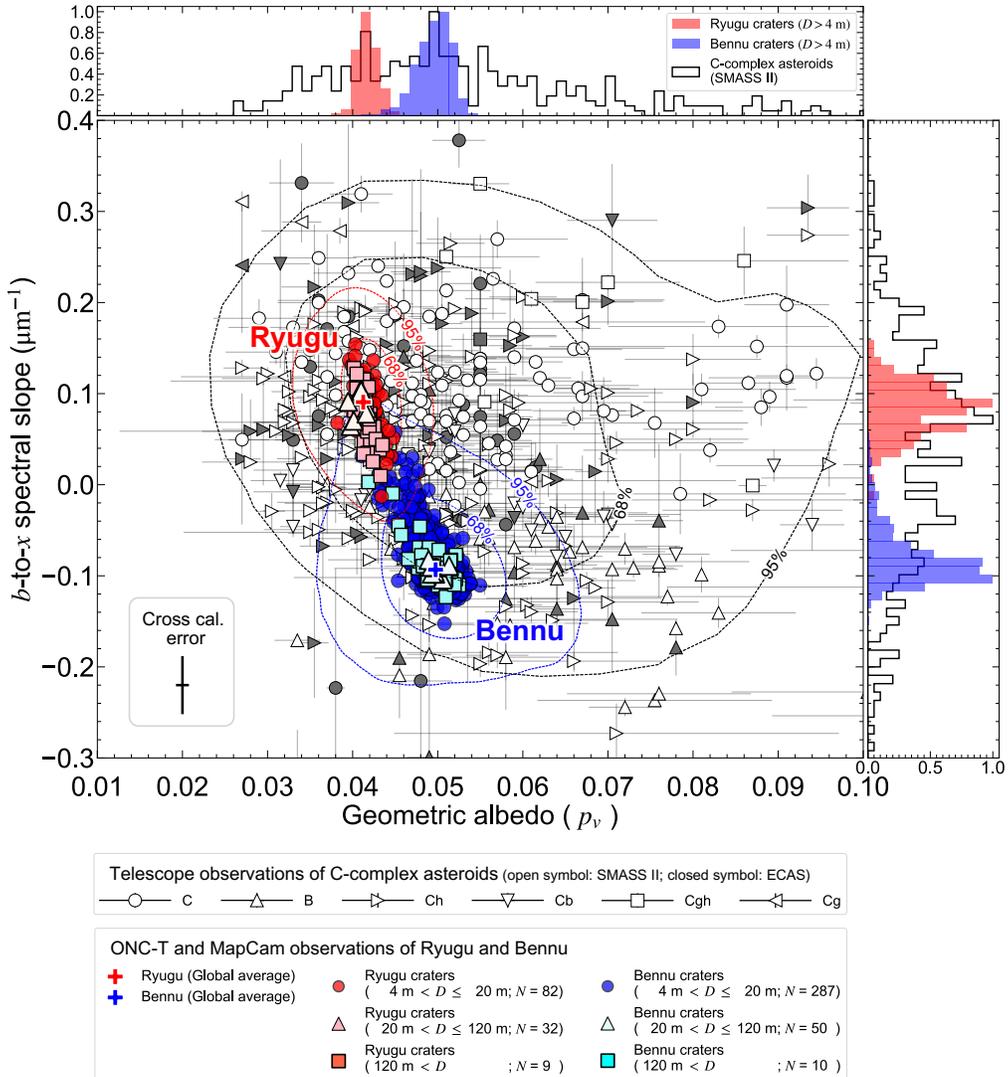

**Figure 13**. Albedo–spectral slope distributions of craters on Ryugu and Bennu compared with those of C-complex asteroids observed by ground-based telescopes (SMASS II and ECAS). The plots for C-complex asteroids are categorized based on their spectral types: C, B, Ch, Cb, Cgh, and Cg (Bus & Binzel, 2002b). The thin error bars of each asteroid data show the measurement errors. The histograms on the top and right show the frequency of crater and asteroid normalized with maximum count. Symbols "+" show the global average of Ryugu and Bennu. The red and blue dashed lines show the 68 and 95% ($1\sigma$ and $2\sigma$) percentiles within the 15ºN to 30ºS region on Ryugu and Bennu observed at 0.3 m/pix (same as Fig. 5c). The dark dashed lines show the 68 and 95% ($1\sigma$ and $2\sigma$) percentiles of C-complex asteroid spectra. The thick black error bar in the lower left corner indicates the magnitude of cross-calibration error between ONC-T and MapCam, demonstrating that the bias between Ryugu and Bennu data is smaller than this range. The cross-calibration error is ±2% for albedo and ±0.03 $\mu m^{-1}$ for *b*-to-*x* spectral slope.

## 6. Conclusion

We compared the optical spectra of Ryugu and Bennu in the *b* (480 nm), *v* (550 nm), *w* (700 nm), and *x* (850 nm) bands after correcting for the imager-to-imager bias using the cross-calibration results obtained in our companion paper. We showed that the global spectrum of Bennu is significantly brighter by 18.0 ± 1.5% at *v* band and bluer by 0.18 ± 0.03 μm$^{-1}$ in the *b*-to-*x* band range than Ryugu. The asteroid-to-asteroid difference is larger than the latitudinal variation within each asteroid. Despite such significant differences between the two asteroids, the spectra of certain surface features are similar. For instance, the reflectance and spectral slope of the Otohime saxum on Ryugu are similar to Bennu's global average.

We found that the spatial scales of spectral heterogeneity differ between Ryugu and Bennu. Boulder-to-boulder variations dominate the spectral slope heterogeneity on Bennu, while on Ryugu, the spectral slope heterogeneity has a greater contribution from larger geologic features (≳10−100 m). The magnitude of spectral slope heterogeneity reverses at ~1 m resolution; heterogeneity is larger for Bennu at scales smaller than 1 m, but Ryugu is more heterogeneous at scales larger than 1 m. More efficient lateral mixing of surface materials by resurfacing processes on Bennu than on Ryugu may account for the homogenization of spectral slopes over longer distances.

We showed that the spectral distributions of craters on Ryugu and Bennu formed two parallel lines with an offset before applying the results of cross calibration. However, the two parallel trend lines converged into one after applying the results of cross calibration; the offset was caused by imager-to-imager bias.

Furthermore, we found that the optical spectra of the freshest craters on Ryugu and Bennu show remarkable similarities; the two are indistinguishable within the uncertainties of cross calibration. The Hokioi crater on Bennu, which contains the Nightingale sample collection site, also has a spectrum indistinguishable from the freshest craters on Ryugu. In addition, we show that the change in spectra from smaller to larger craters, which likely reflects the spectral alterations due to space weathering, occurs in completely opposite directions along a common straight trend line.

These findings suggest that Ryugu and Bennu initially had similar spectra before space weathering and that they evolved in completely opposite directions along the same trend line in the reflectance−spectral slope diagram (i.e., darkening−reddening on Ryugu and brightening−bluing on Bennu), subsequently evolving into asteroids with different spectra (i.e., Cb type for Ryugu and B type for Bennu).

We suggested that the opposite trends of space weathering on Ryugu and Bennu may be caused by the evolution of physical properties, such as the thickness of dust coating and grain size, of the asteroid surface material. For instance, it may be possible that the abundance of fine particles, contributing to darkening and reddening, increased over time on Ryugu, while on Bennu, these particles were preferentially lost, causing the spectra of these asteroids to evolve along the same trend line in opposite directions. Nevertheless, we cannot entirely rule out the possibility that differences in composition, weathering dose, or weathering agent (e.g., solar wind versus micrometeorite) between Ryugu and Bennu led to varying chemical or mineralogical alterations due to space weathering effects.

Ryugu and Bennu demonstrate that space weathering is likely the major cause of their spectral difference (i.e., Cb-type vs. B-type), suggesting that space weathering has likely expanded the spectral slope variation among C-complex asteroids. This further implies that C-complex asteroids may have formed from materials with more uniform spectral slopes than the variation observed today by ground-based telescopes.

**Acknowledgements**

We thank two anonymous reviewers for their helpful comments. We are grateful to the entire OSIRIS-REx and Hayabusa2 teams for making the encounters with Bennu and Ryugu possible. This publication makes use of data products from the Wide-field Infrared Survey Explorer, which is a joint project of the University of California, Los Angeles, and the Jet Propulsion Laboratory/California Institute of Technology, funded by the National Aeronautics and Space Administration. K. Y acknowledges funding from JSPS Fellowship (Grant numbers JP21J20894 and 24KJ2228) and International Graduate Program for Excellence in Earth-Space Science (IGPEES) from the University of Tokyo. E. T. and S. S acknowledge funding by the International Visibility Program of Hayabusa2# project from JAXA. S. S also acknowledges funding by JSPS grants (Grant numbers 22K21344 and 23H00141). This work was part of K. Y.'s doctoral thesis, and we extend our gratitude to the reviewers A. Takigawa, T. Hirata, N. Namiki, M. Ohtake, and Y. Takahashi for their thorough review of the thesis and their valuable comments.

**Data availability statement**

The calibrated (iofL2) MapCam images are available via the Planetary Data System (PDS): https://sbn.psi.edu/pds/resource/orex/. The calibrated (L2d) ONC-T images are also available via PDS: https://sbn.psi.edu/pds/resource/hayabusa2/.

**Supplementary materials**

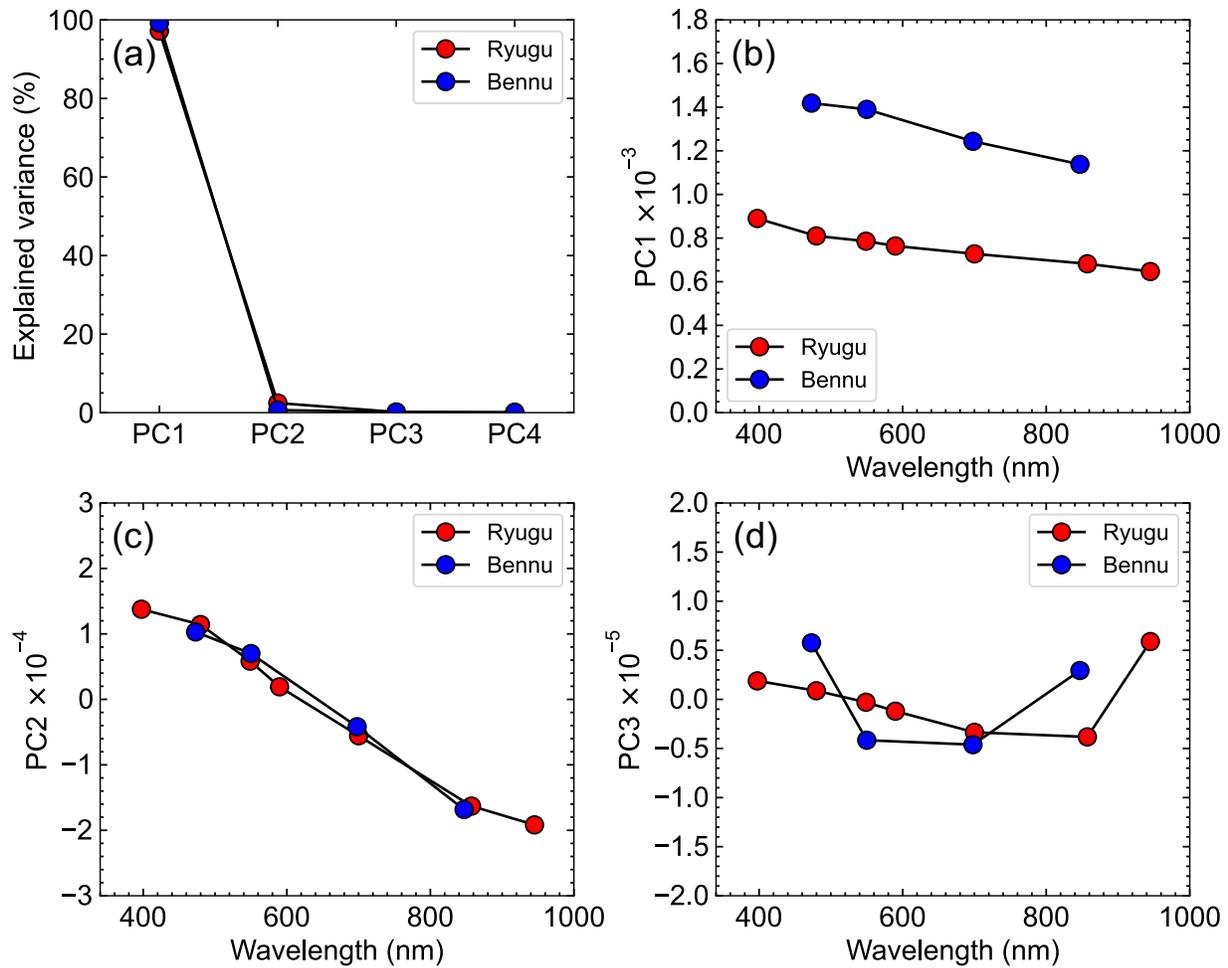

**Figure. S1.** Results of principal component analysis of crater spectra. **(a)** The variance explained by the first four principal components. The first, second, and third principal components are shown in **(b)**, **(c)**, and **(d)**, respectively.

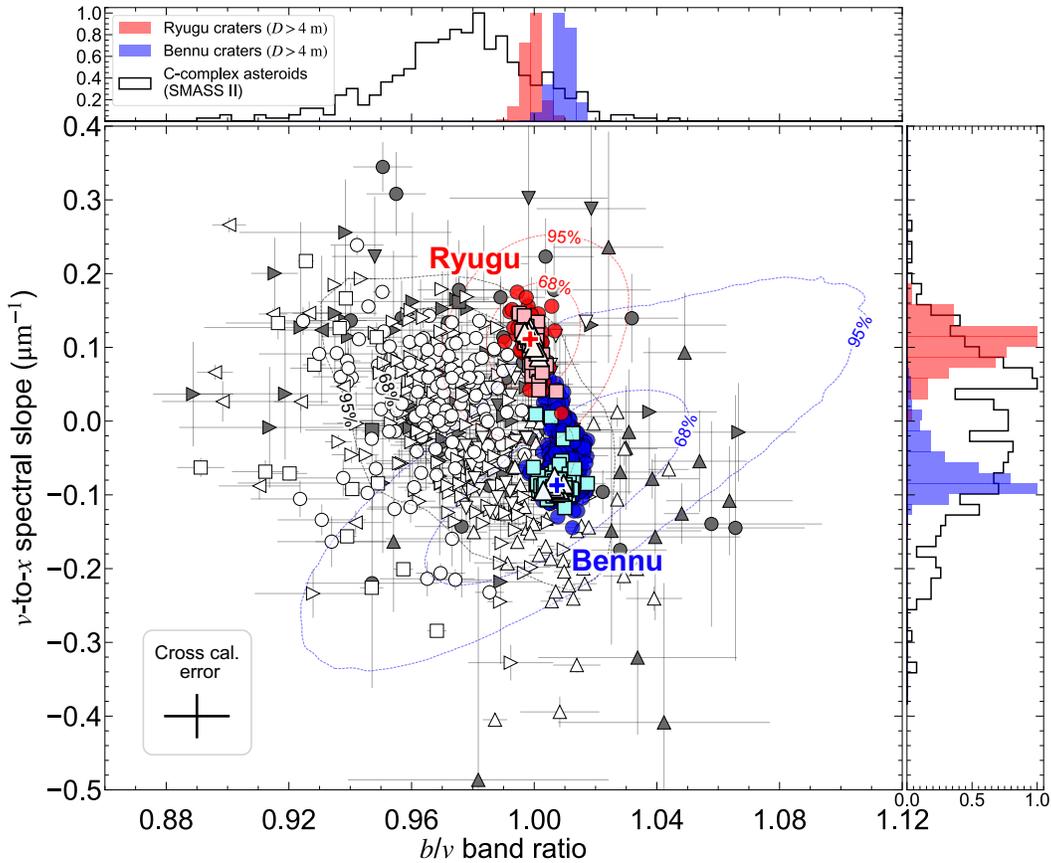
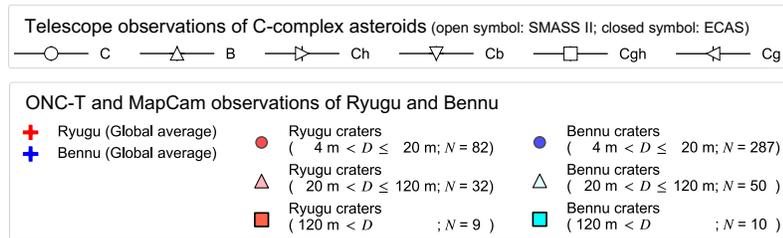

**Figure S2.** The *b*/*v* band ratio vs. *v*-to-*x* band spectral slope distributions of craters on Ryugu and Bennu compared with those of C-complex asteroids observed by ground-based telescopes (SMASS II and ECAS). The plots for C-complex asteroids are categorized based on their spectral types: C, B, Ch, Cb, Cgh, and Cg (Bus & Binzel, 2002). The symbols and notations are the same as Fig. 13. Note that the vertical axis shows the spectral slope in the *b*-to-*x* band range in Fig. 13 but in the *v*-to-*x* band range in Fig. S2. The cross-calibration error is ±1% for band ratio and ±0.03 $\mu m^{-1}$ for *v*-to-*x* spectral slope.

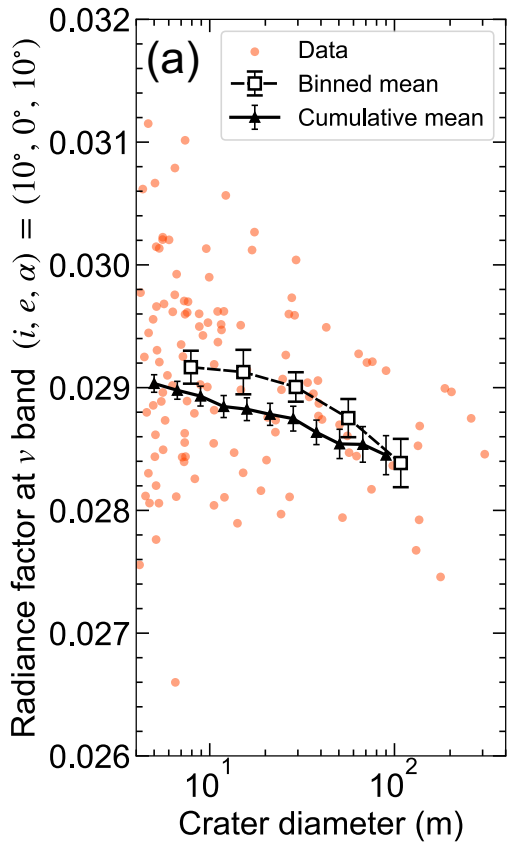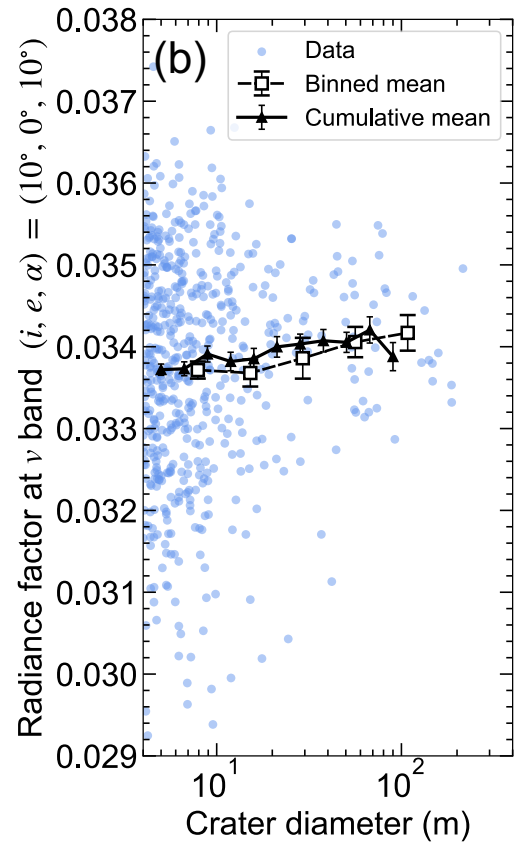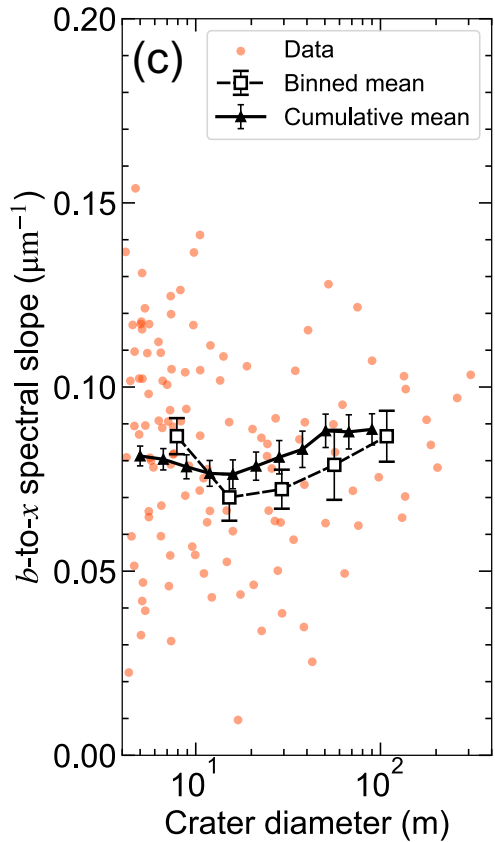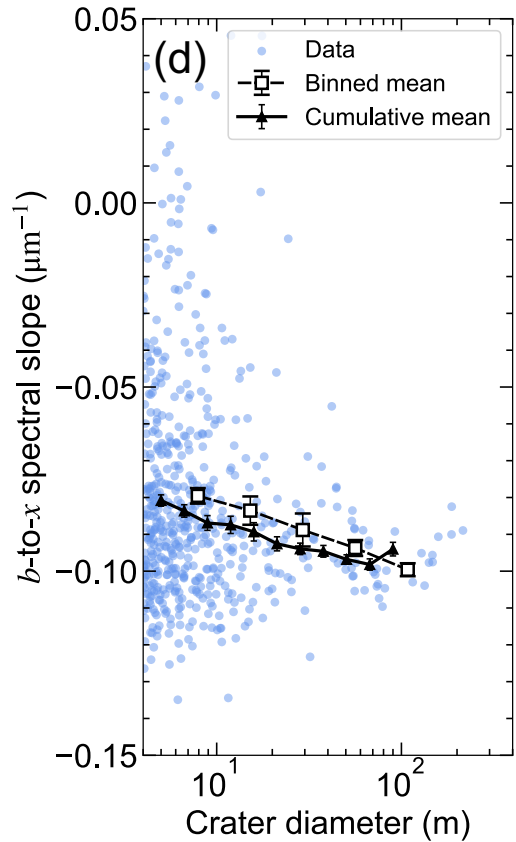

**Figure S3.** Gradual change of crater spectra with its size. The *v*-band reflectance of **(a)** Ryugu and **(b)** Bennu craters and the *b*-to-*x* spectral slope of **(c)** Ryugu and **(d)** Bennu craters are plotted against their diameter. The dashed lines show the average spectra within crater diameter bins evenly spaced on a logarithmic scale. The solid lines show the cumulative mean (i.e., the average of all data larger than the crater diameter on the x-axis). The error bars show the standard error.

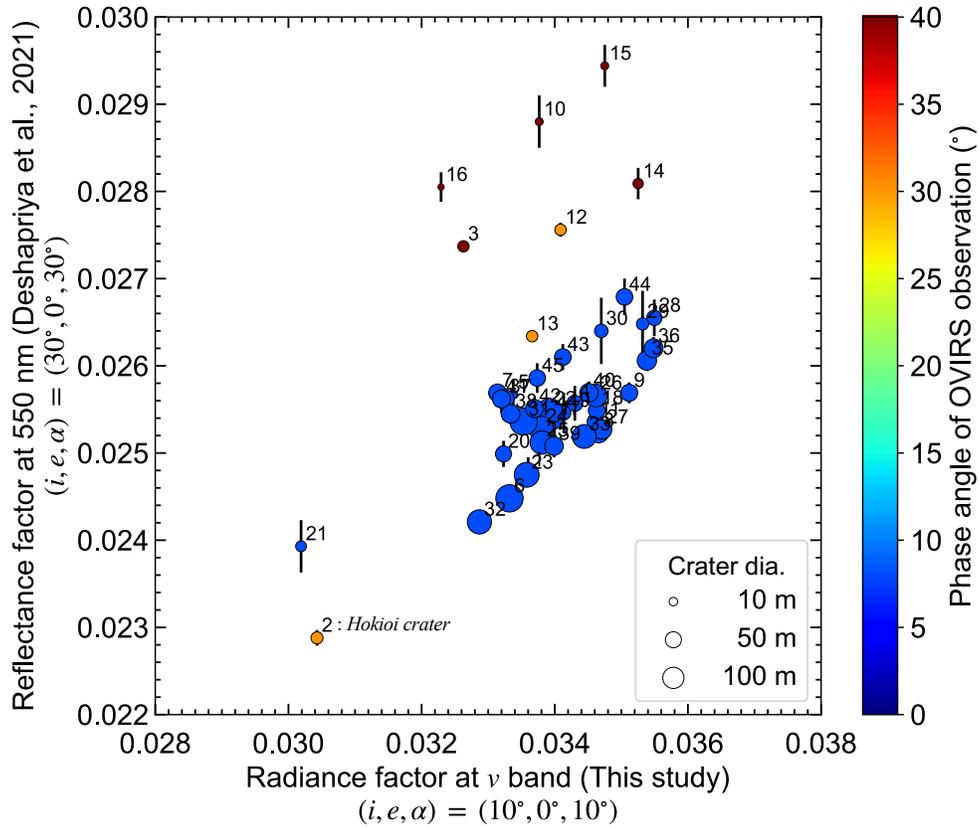

**Figure S4.** Correlation between the crater reflectance obtained in this study based on MapCam data at *v* band ($r'_{norm,v}$) and those reported by Deshapriya et al. (2021) based on OVIRS observations at 550 nm. The colors of each plot show the phase angles during the OVIRS observations. All craters were observed with phase angles of 8° for MapCam data. Each plot is labelled with the crater ID reported in Deshapriya et al. (2021). Error bars show the uncertainties reported in Deshapriya et al. (2021) combining instrumental, calibration, and photometric correction errors.